\documentclass[longauth]{aa} 
\usepackage[varg]{txfonts} 
\usepackage{graphicx}
\usepackage{natbib}
\usepackage{amsmath} 
\bibpunct{(}{)}{;}{a}{}{,} 
\setcitestyle{notesep={ }} 
\usepackage{multirow}
\usepackage{hyperref}
\hypersetup{
colorlinks,
citecolor=blue,
}
\usepackage{xcolor} 

\newcommand{\zspec}{$z_{\rm{spec}}$}
\newcommand{\zphot}{$z_{\rm{phot}}$}
\newcommand{\zmode}{$z_{\rm{mode}}$}
\newcommand{\zmean}{$z_{\rm{mean}}$}
\newcommand{\zmedian}{$z_{\rm{median}}$}
\newcommand{\zrandom}{$z_{\rm{random}}$}
\newcommand{\snmad}{$\sigma_{\rm{NMAD}}$}
\newcommand{\fone}{$f_{\rm{1}}$}
\newcommand{\etafifteen}{$\eta_{\rm{15}}$}
\newcommand{\etathree}{$\eta_{\rm{3}}$}
\newcommand{\pstar}{$P_{\rm{star}}$}
\newcommand{\deltaonetwo}{$\Delta_{1,2}$}
\newcommand{\sdet}{$S_{\rm{det}}$}
\newcommand{\pnoise}{$P_{\rm{noise}}$}

\defcitealias{HC21}{HC21} 

\begin{document}

\title{J-PLUS: Spectral classification and photometric redshifts for 79 million sources in the fourth data release} 
\author{A.~Hern\'an-Caballero\inst{\ref{CEFCA},\ref{UA}} \thanks{email: ahernan@cefca.es}
\and H.~V\'azquez Rami\'o\inst{\ref{CEFCA},\ref{UA}}
\and C.~L\'opez-Sanjuan\inst{\ref{CEFCA},\ref{UA}}
\and D.~Muniesa\inst{\ref{CEFCA}}
\and T.~Civera\inst{\ref{CEFCA}}
\and J.~A.~Fern\'andez-Ontiveros\inst{\ref{CEFCA},\ref{UA}}
\and H.~Dom\'inguez S\'anchez\inst{\ref{IFCA}}
\and A.~del Pino\inst{\ref{IAA}}
\and J.~A.~L.~Aguerri\inst{\ref{IAC},\ref{ULL}}
\and S.~Zarattini\inst{\ref{CEFCA},\ref{UA}}
\and V.~Marra\inst{\ref{UFOP},\ref{INAF},\ref{IFPU}}
\and A.~J.~Cenarro\inst{\ref{CEFCA},\ref{UA}}
\and D.~Crist\'obal-Hornillos\inst{\ref{CEFCA}}
\and C.~Hern\'andez-Monteagudo\inst{\ref{IAC},\ref{ULL}}
\and A.~Mar\'{\i}n-Franch\inst{\ref{CEFCA},\ref{UA}}
\and M.~Moles\inst{\ref{CEFCA}}
\and J.~Varela\inst{\ref{CEFCA}}
\and J.~Alcaniz\inst{\ref{ON}}
\and R.~A.~Dupke\inst{\ref{ON},\ref{MU}}
\and A.~Ederoclite\inst{\ref{CEFCA}}
\and L.~Sodr\'e Jr.\inst{\ref{USP}}
\and R.~E.~Angulo\inst{\ref{DIPC},\ref{ikerbasque}}
\and J.~Zaragoza-Cardiel\inst{\ref{CEFCA},\ref{UA}}
}

\institute{
Centro de Estudios de F\'{\i}sica del Cosmos de Arag\'on (CEFCA), Plaza San Juan 1,
44001 Teruel, Spain\label{CEFCA}
\and Unidad Asociada CEFCA-IAA, CEFCA, Unidad Asociada al CSIC por el IAA y el IFCA,
Plaza San Juan 1, 44001 Teruel, Spain\label{UA}
\and Instituto de F\'isica de Cantabria, Av. de los Castros, 39005 Santander, Cantabria, Spain\label{IFCA}
\and Insitituto de Astrof\'isica de Andaluc\'ia (CSIC), PO Box 3004, 18080 Granada, Spain\label{IAA}
\and Departamento de Física, Universidade Federal de Ouro Preto, 35400-000, Ouro Preto, MG, Brazil\label{UFOP}
\and INAF -- Osservatorio Astronomico di Trieste, via Tiepolo 11, 34131 Trieste, Italy\label{INAF}
\and IFPU -- Institute for Fundamental Physics of the Universe, via Beirut 2, 34151, Trieste, Italy\label{IFPU}
\and Instituto de Astrof\'{\i}sica de Canarias, La Laguna, 38205, Tenerife, Spain\label{IAC}
\and Departamento de Astrof\'{\i}sica, Universidad de La Laguna, 38206, Tenerife,
Spain\label{ULL}
\and Observat\'orio Nacional - MCTI (ON), Rua Gal. Jos\'e Cristino 77, S\~ao Crist\'ov\~ao,
20921-400 Rio de Janeiro, Brazil\label{ON}
\and University of Michigan, Department of Astronomy, 1085 South University Ave., Ann
Arbor, MI 48109, USA\label{MU}
\and Instituto de Astronomia, Geof\'{\i}sica e Ci\^encias Atmosf\'ericas, Universidade de
S\~ao Paulo, 05508-090 S\~ao Paulo, Brazil\label{USP}
\and Donostia International Physics Centre (DIPC), Paseo Manuel de Lardizabal 4, 20018
Donostia-San Sebastián, Spain\label{DIPC}
\and IKERBASQUE, Basque Foundation for Science, 48013, Bilbao, Spain\label{ikerbasque}
}

\date{Accepted ........ Received ........;}

\abstract {
We present spectral classifications and photometric redshifts for 79.2 million sources up to an \textit{r}-band magnitude of 22 in Data Release 4 of the Javalambre Photometric Local Universe Survey (J-PLUS). Leveraging the 12-band J-PLUS filter system, we compare a template-fitting approach (\textsc{LePhare}) against \textsc{LeMoNNADE}, a morphology-blind machine learning pipeline that uses spectral mixing augmentation to overcome training set limitations. \textsc{LeMoNNADE} consistently outperforms template fitting in precision, robust scatter, and outlier rates. Including WISE infrared photometry breaks optical degeneracies between stars and quasars, reducing the catastrophic outlier rate for quasars from $\sim$40\% to $\sim$23\% and constraining systemic redshift bias to $\pm$1\% up to \textit{z} = 4. We find \textsc{LeMoNNADE} is also less susceptible to redshift aliasing, particularly when adopting the probability density function median. 
Because the spectroscopic training samples severely under-represent stars, we apply an Expectation-Maximization Bayesian calibration to recover unbiased class probabilities for the magnitude-limited sample. This reveals that extragalactic counts agree with the literature down to the \textit{r} $\sim$ 20.5 completeness limit. 
The inferred redshift distribution for $r < 21$ extragalactic sources peaks at $z \sim 0.3$, showing broad agreement with existing literature up to $z \sim 0.6$.
The resulting catalogues represent a significant milestone for local Universe science, offering probabilistically calibrated classifications and distances while explicitly characterising faint-end limits and contamination.
}


\titlerunning{Photo-z for J-PLUS DR4}
\authorrunning{A. Hern\'an-Caballero et al.}

\maketitle

\section{Introduction}

The Javalambre Photometric Local Universe Survey \citep[J-PLUS;][]{Cenarro2019} occupies a unique niche among modern large-area imaging programmes. Its optical filter system, composed of 12 passbands, bridges the gap between traditional broad-band surveys \citep[e.g., SDSS, Pan-STARRS;][]{York00, Chambers2016} and wide-area narrow-band surveys such as J-PAS \citep{Benitez2014, Bonoli2021}. By combining the standard broad $ugriz$ bands with seven narrow bands strategically positioned to capture key diagnostic spectral features at $z \sim 0$, J-PLUS provides an unprecedented low-resolution spectral sampling for millions of sources in the local Universe.

With Data Release 4 (DR4; V\'azquez Rami\'o et al. in prep.), J-PLUS reaches a significant milestone, not only in terms of sky coverage (spanning $\sim$5000 deg$^2$ observed, with an effective area of 4437.5 deg$^2$) but also in the maturity of its data reduction and calibration pipeline. These advancements have resulted in a high-precision absolute photometric calibration, achieving rms errors of $\lesssim 10$ mmag across the entire survey footprint (V\'azquez Rami\'o et al. in prep.).

Such photometric homogeneity, combined with the availability of spectroscopic redshifts and classifications for millions of J-PLUS sources, presents a unique opportunity: it allows for the training of large machine learning models for photometric redshift (photo-$z$) estimation and spectral classification directly on the observed J-PLUS photometry, effectively bypassing the need for assumptions regarding the intrinsic spectral energy distributions (SEDs) of the sources.

In this work, we present the photometric redshift and spectral classification catalogues for J-PLUS DR4. These catalogues include $\sim$79 million unique sources up to a magnitude limit of $r < 22$. To maximise the reliability and versatility of our estimates, we employ a dual approach to redshift estimation. First, we use {\sc LePhare} \citep{Arnouts2011, Ilbert2006}, a widely used template-fitting code that we configure with an optimised template set and priors derived specifically for J-PLUS. In this configuration, the redshift solutions are restricted to galaxies. Second, we introduce results from {\sc LeMoNNADE} (Learning Model with Neural Network for Adaptive Distance Estimation), a novel machine learning pipeline tailored for large-area multi-band imaging surveys (Hern\'an-Caballero et al. in prep.). 

Unlike machine-learning tools previously applied to source classification in J-PLUS DR3, such as BANNJOS \citep{delPino2024} and XGBoost \citep{vonMarttens2024}, {\sc LeMoNNADE} deliberately excludes morphological information. By relying exclusively on aperture-corrected photometry, the model ensures robustness against systematics introduced by variations in the point spread function (PSF). Furthermore, to mitigate the ambiguity inherent in the spectroscopic classification of galaxies hosting active galactic nuclei at low and intermediate redshifts, galaxies and quasars (QSOs) are grouped into a single ``extragalactic'' class. The {\sc LeMoNNADE} photo-$z$ model is trained on a comprehensive sample of $\sim$2 million extragalactic objects. To overcome the limitations of the training set and prevent overfitting, the pipeline introduces spectral mixing augmentation, which effectively interpolates across gaps in the colour-redshift parameter space. Additionally, the model uses multiple inference iterations with adaptive variance scaling to produce highly realistic probability distribution functions (PDFs) that accurately represent both epistemic and stochastic uncertainties.

The paper is structured as follows. Section \ref{sec:data} describes the J-PLUS DR4 dataset, the ancillary photometric data, and the spectroscopic training samples. We detail the configuration of the template-fitting method in Section \ref{sec:lephare} and the machine-learning methodology in Section \ref{sec:lemonnade}. In Section \ref{sec:results}, we evaluate the accuracy of the photo-$z$ and classification performance against spectroscopic control samples. The magnitude and redshift distribution of the full J-PLUS source population is analised in Section \ref{sec:jplus-population}. Finally, we summarise our conclusions in Section \ref{sec:conclusions}. Throughout this paper, all magnitudes are reported in the AB system.

\section{Data}
\label{sec:data}

\subsection{J-PLUS photometry}

\begin{table}
\caption{Properties of J-PLUS coadded images\label{table:jplus-tiles}}
\centering
\begin{tabular}{c c r c c c}
\hline\hline
Filter & $\lambda_{\rm{central}}$ & \multicolumn{1}{c}{$\Delta\lambda$} & FWHM  & Depth 5-$\sigma$ & $m_{50}$ \\
       & [nm] & [nm] & [\arcsec] & [mag] & [mag] \\
\hline
$u$ & 348.5 & 45.8 & 1.35 & 20.93 & 20.97 \\
$g$ & 475.0 & 125.0 & 1.28 & 21.89 & 21.96 \\
$r$ & 625.0 & 125.0 & 1.20 & 21.94 & 22.04 \\
$i$ & 772.5 & 125.0 & 1.14 & 21.66 & 21.76 \\
$z$ & 915.0 & 130.0 & 1.12 & 20.77 & 20.88 \\
\textit{J0378} & 378.5 & 13.8 & 1.35 & 20.88 & 20.93 \\
\textit{J0395} & 395.0 & 7.0 & 1.33 & 20.89 & 20.94 \\
\textit{J0410} & 410.0 & 17.0 & 1.32 & 21.16 & 21.22 \\
\textit{J0430} & 430.0 & 17.0 & 1.29 & 21.19 & 21.26 \\
\textit{J0515} & 515.0 & 17.0 & 1.26 & 21.20 & 21.29 \\
\textit{J0660} & 660.0 & 10.0 & 1.19 & 21.25 & 21.37 \\
\textit{J0861} & 861.0 & 36.0 & 1.11 & 20.55 & 20.72 \\
\hline
\end{tabular}
\begin{flushleft}
\footnotesize
Notes: $\lambda_{\rm{central}}$: Central wavelength; $\Delta\lambda$: Filter width; 
FWHM: Median PSF FWHM of image; Depth 5-$\sigma$: Median magnitude at SNR=5 in 3 arcsec aperture; 
$m_{50}$: Median magnitude at 50\% detection rate for point sources.
\end{flushleft}
\end{table}

J-PLUS is being conducted at the Observatorio Astrof\'{\i}sico de Javalambre (OAJ; \citealt{oaj}) with the 83\,cm Javalambre Auxiliary Survey Telescope (JAST80) and the 9200 $\times$ 9200 pixel camera T80Cam \citep{t80cam}, providing a 2\,deg$^2$ field of view per pointing. 
The J-PLUS DR4 footprint covers an observed area of 4950.7 deg$^2$, resulting in an effective area of 4437.5 deg$^2$ after accounting for masked regions and the area of overlap between adjacent tiles. All 2653 tiles in DR4 have been observed in the 12 J-PLUS photometric bands. The median Point Spread Function (PSF) Full Width at Half Maximum (FWHM) and the limiting depths for each band are summarised in Table \ref{table:jplus-tiles}.

The data reduction and calibration for J-PLUS DR4 is thoroughly described in V\'azquez Rami\'o et al. (in prep.). A brief summary is described here.
Reduction of single exposures included bias subtraction, flat-fielding, and illumination correction. In the $z$ band, a fringing correction was also applied. The astrometric calibration was performed using {\sc Scamp} \citep{Bertin2006}, with Gaia DR3 as the reference catalogue. 
The photometric calibration of single exposures was processed in four steps to translate instrumental magnitudes to the AB system, achieving $\sim 1\%$ accuracy and precision across all 12 passbands \citep{LopezSanjuan2024}. 
First, single exposures were calibrated by comparing instrumental magnitudes of main-sequence stars to synthetic photometry in the J-PLUS bands obtained from the BEst STar database \citep[BEST;][]{Xiao2026}, which is derived from calibrated Gaia DR3 BP/RP low-resolution spectra. Calibrated single exposures are normalised to an arbitrary zero point of AB magnitude 25.
Second, the photometry was spatially flattened across the image using the residuals of the initial calibration. This illumination correction accounts for both magnitude and colour terms to resolve spatial variations of the zero-point across the focal plane.
Third, the absolute AB colour scale was established using the theoretical locus of 242 nearby ($d < 100$ pc) white dwarfs with negligible interstellar extinction. Finally, the absolute flux scale was anchored to Pan-STARRS1 photometry using the $r$ band.
Subsequently, the individual exposures were co-added using {\sc Swarp} \citep{Bertin2002}, with all images resampled to a pixel scale of 0.45\arcsec to minimise aliasing.

Source detection and extraction were performed on the co-added images using {\sc SExtractor} \citep{Bertin1996}. We employed a dual-mode photometry strategy, where the detection image is used to define the apertures and the measurement is performed on the corresponding co-added image for each band. The $r$ band served as the reference detection image. The resulting dual-mode catalogue for the entire J-PLUS DR4 comprises 96.9 million unique sources, after accounting for duplicates in the overlapping regions between adjacent tiles.

For the purposes of photometric redshift estimation and spectral classification, we select sources from the dual-mode catalogues that satisfy the following criteria: a) magnitude $r < 22$ in the \texttt{AUTO} aperture; b) \texttt{FLAGS} $\le 3$ in the $g$, $r$, and $i$ bands (which removes saturated stars and objects affected by severe artefacts). The final cleaned sample contains 78.6 million unique sources.

Although the sample selection uses \texttt{AUTO} magnitudes for completeness, we adopt the 3\arcsec\ diameter photometry, corrected for aperture effects \citep[\texttt{APER\_COR\_3\_0};][]{LopezSanjuan2022} for the computation of photo-$z$ and spectral classification. As detailed in V\'azquez Rami\'o et al. (in prep.), this aperture yields the most accurate colour measurements. To preserve this colour fidelity while recovering the total flux (which is needed to apply the correct luminosity prior on extended sources) we scale the \texttt{APER\_COR\_3\_0} SED of each source by the ratio of the \texttt{AUTO} to \texttt{APER\_COR\_3\_0} flux in the $r$ band.

The photometry supplied to the {\sc LePhare} pipeline undergoes three additional pre-processing steps:
\begin{enumerate}
    \item Correction for Galactic dust extinction: We calculate the attenuation $a_\lambda$ in each band using the Bayestar17 dust maps \citep{Green2018}, adopting the extinction coefficients $k_\lambda$ from \citet{Whitten2019} and the extinction law of \citet{Schlafly2016}.
    \item Application of global photometric calibration offsets: These offsets correct for small discrepancies between the nominal and effective transmission of the J-PLUS filters, determined via the `recalibration' procedure described in \citet{HC21}.
    \item Error floor injection: A systematic uncertainty of 0.01 magnitudes is added in quadrature to the photometric errors in all bands to account for residual calibration uncertainties.
\end{enumerate}

We emphasise that these pre-processing steps are not applied to the input photometry for {\sc LeMoNNADE}. First, global calibration offsets do not affect machine learning models trained directly on the uncorrected observed data. Second, applying 2D Galactic dust maps is inappropriate for stars, as it assumes the total line-of-sight dust column and overestimates their attenuation.

\subsection{WISE photometry}

As discussed in Section \ref{sec:accuracy-binclass}, the J-PLUS photometry alone (hereafter the \texttt{jplusonly} dataset) presents challenges for {\sc LeMoNNADE} in separating faint blue stars from QSOs, and faint red stars from galaxies. To mitigate this degeneracy, we incorporate infrared photometry in the W1 and W2 bands of WISE from the unWISE catalogue \citep[][; hereafter the \texttt{jplus+wise} dataset]{Schlafly2019}.

We performed a positional cross-match between each J-PLUS tile catalogue and the overlapping unWISE tiles using a search radius of 1.5\arcsec. We adopted the local-background-subtracted PSF photometry provided by unWISE (columns \texttt{fluxlbs} and \texttt{dfluxlbs}). The unWISE catalogue only provides fluxes for $5\sigma$ detections. The fraction of $r < 22$ J-PLUS sources with detections in the WISE W1 and W2 bands is 69.2\% and 51.5\%, respectively.
For J-PLUS sources lacking WISE detections, we estimate $5\sigma$ upper limits based on the local background noise. We model these limits as a function of the ecliptic latitude, $\beta$, by taking half the average flux of $10\sigma$ sources in bins of |$\beta$|. A linear regression of the background dependence yields the following upper limits:
\begin{eqnarray}
f_{5\sigma,W1} / \mu\mathrm{Jy} &\approx& 12.5 - 0.075 \left| \beta / 1^{\circ} \right|, \\
f_{5\sigma,W2} / \mu\mathrm{Jy} &\approx& 27.5 - 0.150 \left| \beta / 1^{\circ} \right|
\end{eqnarray}

\subsection{Spectroscopic redshifts}
\label{sec:specz}

\begin{figure}
\begin{center}
\includegraphics[width=8.4cm]{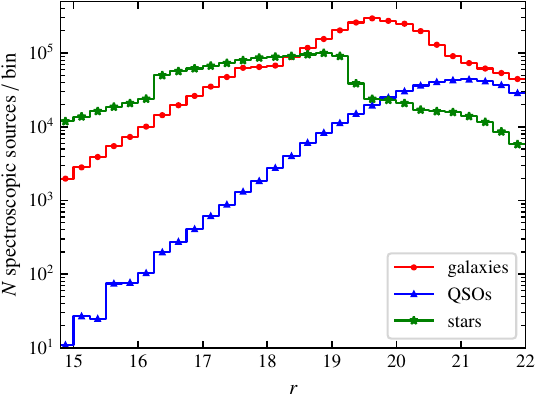} 
\end{center}
\vspace{-0.4cm}
\caption[]{Magnitude distribution for galaxies, quasars and stars in the J-PLUS DR4 subsample with spectroscopy. The notable increase in star counts in the range $16.2 < r < 19.2$ is due to the selection criteria of the DESI Milky Way Survey \citep{Cooper2023}.\label{fig:mag-distrib}}
\end{figure}

To build the training and validation samples for the machine learning algorithms, and to assess the accuracy of the template-fitting results, we compiled a spectroscopic catalogue from four major surveys: the Dark Energy Spectroscopic Instrument (DESI) Data Release 1 \citep{DESI-DR1}, the Sloan Digital Sky Survey (SDSS) Data Release 17 \citep{SDSS-DR17}, HectoMAP Data Release 2 \citep{Sohn2023}, and the Hobby-Eberly Telescope Dark Energy Experiment (HETDEX) Public Source Catalog 1 \citep{MentuchCooper2023}.
Initially, we selected all the spectroscopic sources located within the J-PLUS DR4 footprint.
To ensure the reliability of the redshift ground truth, we excluded sources with quality flags indicating potential measurement failures (e.g. \texttt{zWarning} $> 0$ in DESI or SDSS).
The raw counts for each survey prior to duplicate removal are 5.37 million (DESI), 1.38 million (SDSS), 101,949 (HectoMAP), and 65,302 (HETDEX).

Given the significant spatial overlap between these surveys, we performed an internal cross-match using a search radius of $1\arcsec$.
For sources identified in multiple catalogues, we adopted the redshift and spectral classification from the survey with the highest priority, defined by the following hierarchy: DESI $>$ SDSS $>$ HectoMAP $>$ HETDEX.
The resulting combined spectroscopic catalogue contains 6,499,773 unique sources.

We subsequently cross-matched this combined catalogue with the cleaned J-PLUS DR4 photometric catalogue (applying the cuts $r < 22$ and \texttt{FLAGS} $\le 3$ described in Sect.~\ref{sec:data}) using a matching radius of $1\arcsec$.
This yielded a final sample of 4,153,894 unique J-PLUS sources with reliable spectroscopic redshifts. The distribution of $r$-band magnitude for stars, galaxies and QSOs in this sample is shown in Figure \ref{fig:mag-distrib}.

For the training of the \textsc{LeMoNNADE} models, this dataset was randomly partitioned into three subsets: a training set (96\%) used to optimise the neural network weights, a validation set (2\%) monitored to detect overfitting and tune hyperparameters, and a test set (2\%) strictly reserved for generating the final performance statistics presented in Sect.~\ref{sec:results}. While the validation and test sets are unusually small relative to the training set, at $\sim$83,000 sources each they are large enough to be representative of the entire spectroscopic sample.

The selection criteria of the spectroscopic surveys imply that these spectroscopic samples are far from representative of the population of $r$$<$22 J-PLUS sources. In particular, faint stars are strongly underrepresented. In Sect.~\ref{sec:jplus-population} we discuss the bias that this introduces in the star/extragalactic classification and present our mitigation strategy.

\section{Photo-z with {\sc LePhare}}
\label{sec:lephare}

The SED-fitting variant of J-PLUS photo-$z$ are computed using a modified version of the code {\sc LePhare} \citep{Arnouts2011, Ilbert2006} customised to address the specific needs of the Javalambre surveys. Notably, the maximum number of photometric bands has been increased to 64 to accommodate the J-PAS filter system, and several non-essential outputs have been disabled to enhance computational efficiency.

We employ a specific configuration using custom spectral templates and a redshift prior tailored to the survey depth and filter system. The redshift probability distribution function (PDF) is sampled in the range $0 \leq z \leq 1$ with a constant step size of $\delta z = 0.005$.

\subsection{Templates}

The template library consists of 30 synthetic stellar population models generated with {\sc CIGALE} \citep{Boquien2019}, selected from a parent sample of 455 candidates following the optimization procedure described in \citet{HC21}. 
During the fitting process, {\sc LePhare} modifies the input templates by adding internal extinction following the \citet{Calzetti2000} attenuation law. We allow for three discrete values of colour excess: $E(B-V) = 0.0, 0.2,$ and $0.4$ mag.

Although \textsc{LePhare} includes libraries for stars and quasars, we follow the strategy of \citet{HC21} and configure the code to use exclusively the galaxy templates described above.
Consequently, the resulting probability distribution functions must be interpreted as conditional probabilities, valid under the assumption that the source is a galaxy.

\subsection{Redshift prior}

The default Bayesian prior in {\sc LePhare} is an analytical model $P(z|m_I,T)$ calibrated to reproduce the redshift distribution of the VIMOS VLT Deep Survey \citep[VVDS;][]{LeFevre2005}. This model relies on the $I$-band magnitude ($m_I$) and broad spectral type ($T$), where the latter is classified as early, intermediate, or late based on the rest-frame $B-I$ colour.

For J-PLUS DR4, we implemented a custom prior, \texttt{nz\_jplus\_dr4}, adapted to use the J-PLUS $i$ band and the restframe $g-i$ colour. We derived a linear transformation between the rest-frame $B-I$ and $g-i$ colours using synthetic photometry of the templates. Consequently, the spectral type classification ranges were updated as follows: early-type ($g-i > 1.0$), intermediate ($0.75 < g-i < 1.0$), and late-type ($g-i < 0.75$).

In the standard {\sc LePhare} implementation, the empirical prior is replaced by a step function ($P(z)=1$ for $z<1$, $0$ otherwise) for sources brighter than $i=20$. While this approach is suitable for deep surveys where the high S/N of the photometry renders the prior less critical, we found that for J-PLUS, such a flat prior resulted in a high rate of catastrophic outliers. To mitigate this, we extrapolated the analytical VVDS model to brighter magnitudes.
However, for sources brighter than $i \sim 16$, the extrapolation of the analytical model yields unphysical results (e.g. negative redshifts). Therefore, for sources with $i<16.5$, we replace the analytical prior with the function:
\begin{equation}
    P(z|m_i, T) = \exp\left[-2 \left(\frac{z}{z_{\rm max}(m_i,T)}\right)^2\right]
\end{equation}
where $z_{\rm max}$ is the highest redshift consistent with the observed $i$-band magnitude and a spectral-type dependent threshold in the absolute $i$-band luminosity. We adopt absolute magnitude limits of $M_i = -25$, $-23$, and $-21$ for early, intermediate, and late-type galaxies, respectively.

\subsection{Post-processing of {\sc LePhare} results}
\label{sec:lephare-post}

The probability distribution functions (PDFs) generated by {\sc LePhare} undergo a contrast correction procedure following the methodology described in \citet{HC21}. This step ensures that the PDFs accurately reflect the statistical uncertainties, preventing under- or over-confidence in the redshift estimates. The correction involves two operations: first, the raw PDF is broadened via convolution with a Lorentzian kernel of half-width at half-maximum $\gamma$. Second, the contrast between peaks and valleys is adjusted using the transformation $P'(z) = P(z)^{1/\alpha}$. The parameters $\alpha$ and $\gamma$ are optimised using the spectroscopic training sample. 

Given the immense data volume of J-PLUS DR4, providing full-resolution PDFs is prohibitive. We therefore compress the PDFs using \textsc{ColdPress} \citep{Hernan-Caballero2025}, a Python module developed specifically for this task. \textsc{ColdPress} encodes the probability distributions by storing the redshifts corresponding to specific quantiles of the cumulative distribution function (CDF), rather than sampling the probability density on a fixed grid. This approach achieves high compression ratios while preserving information density, as the PDF sampling naturally becomes finer in regions of higher probability density. 

While {\sc LePhare} provides several point estimates for the redshift and confidence intervals, for consistency we recompute them (also with \textsc{ColdPress}) from the contrast-corrected and compressed PDFs. Specifically, we compute the mode (\zmode), mean (\zmean), median (\zmedian), and a random draw from the distribution (\zrandom), along with their associated \texttt{odds} parameter\footnote{We define the \texttt{odds} of a point estimate $z_X$ as the integral of $P(z)$ in the interval [$z_X$ - 0.03(1+$z_X$), $z_X$ + 0.03(1+$z_X$)].} and various confidence intervals.

\section{{\sc LeMoNNADE}}
\label{sec:lemonnade}

The Learning Model with Neural Network for Adaptive Distance Estimation ({\sc LeMoNNADE}; Hern\'an-Caballero et al., in prep.) is a new machine-learning pipeline designed for spectral classification and photometric redshift estimation in large-area imaging surveys. In this section, we outline its core principles and the configuration adopted for J-PLUS DR4, deferring a comprehensive description to Hern\'an-Caballero et al. (in prep.).

\subsection{Network Architecture}

{\sc LeMoNNADE} deliberately employs a simple, fully connected neural network rather than specialised architectures. This choice is physically motivated: estimating redshift from multi-band photometry is fundamentally a direct, non-linear mapping from a low-dimensional colour space to a distance. Avoiding restrictive geometric priors (such as those imposed by a Convolutional Neural Network) ensures the model architecture does not bias this mapping. 

The input layer contains $2(N_{\rm bands}-1)$ neurons and is followed by 12 hidden layers (with widths decreasing from 400 neurons in the first layers to 128 in the final ones) using LeakyReLU activation.
The output layer and loss function depend on the operational mode. In photo-$z$ mode, the network predicts a full PDF modelled as a Gaussian Mixture Model (GMM) with three components. This continuous representation eliminates the need for a pre-defined redshift grid, which is particularly important given that a single model is trained simultaneously on galaxies and QSOs (which at the depth of J-PLUS span vastly different redshift ranges). 
The network is trained by minimising the loss function:
\begin{equation}
    \mathcal{L} = -\log(P_{z_{\rm spec}} + P_{\rm floor})
\end{equation}
where $P_{z_{\rm spec}}$ is the probability density integrated over the interval $[z_{\rm spec} - \delta z, z_{\rm spec} + \delta z]$, with $\delta z = z_{\rm scale}(1+z_{\rm spec})$. The hyperparameter $z_{\rm scale}$ determines the magnitude of the redshift errors to which the gradient is most sensitive. The term $P_{\rm floor}$ accounts for the non-zero probability of incorrect spectroscopic labels or severe photometric artefacts in the training data, preventing such outliers from dominating the loss gradient. We obtain the best results for J-PLUS using $z_{\rm scale}$ = 0.005 and $P_{\rm floor}$ = 0.01.
In binary classification mode, a single-neuron output layer with a sigmoid activation function predicts \pstar, the probability that a source is a Galactic star, optimizing the binary cross-entropy loss.

Notably, {\sc LeMoNNADE} departs from standard practice by not using dropout regularization. Because the photometric bands lack redundancy, masking inputs destroys essential spectral information. Furthermore, since the underlying mapping from colour to redshift is deterministic, applying dropout within the hidden layers forces the network to build internal redundancy that decreases the capacity and predictive power of the model. Instead, {\sc LeMoNNADE} relies entirely on the data augmentation strategy (Sect. \ref{sec:augmentation}) to prevent overfitting.

\subsection{Data Ingestion}

Another distinctive feature of {\sc LeMoNNADE} is its exclusive reliance on the photometric spectral energy distribution (SED). By deliberately excluding morphological information, the model remains robust against systematic effects arising from variations in the PSF of the observations.  

During pre-processing, after invalid fluxes are flagged and imputed via interpolation, both the flux densities and their uncertainties are normalized by the flux of the source in the reference band (the $r$ band for J-PLUS). This normalization removes information regarding the apparent brightness of the source, preventing the network from learning spurious correlations between magnitude and redshift (or object class) that originate in the selection functions of the spectroscopic training samples and do not apply to the full J-PLUS sample.

\subsection{Data Augmentation}
\label{sec:augmentation}

To prevent the unregularized network from overfitting on its highly dimensional parameter space, {\sc LeMoNNADE} implements `Spectral Mixing Augmentation' (mixaugmentation). 
The sources of the training sample are grouped by redshift, and new synthetic SEDs are generated every few epochs as linear combinations (with random coefficients) of the normalized SEDs from pairs of sources at the same redshift. 

Mixaugmentation effectively interpolates between discrete points in the colour space, providing physically plausible SEDs that cover gaps in the training data. To simulate varying signal-to-noise ratios, Gaussian noise is subsequently injected into these synthetic SEDs. This process increases the effective size of the training sample by about two orders of magnitude and makes overfitting negligible (same loss for the validation sample and the original sources in the training sample).

\subsection{Treatment of Uncertainties}

The GMM output by the network captures the epistemic uncertainty of the model (e.g., colour-redshift degeneracies). However, it does not fully capture the stochastic uncertainty arising from noise in the input photometry. To account for this, {\sc LeMoNNADE} performs 100 inference iterations per source, perturbing the input fluxes in each iteration assuming a normal distribution based on their nominal errors. 

Photometric errors in the training data cause the model to learn a smoothed colour-redshift mapping. Therefore, explicitly perturbing the inputs at face value would double-count this noise, resulting in underconfident (overly broad) PDFs. {\sc LeMoNNADE} resolves this via an adaptive variance scaling mechanism, empirically calibrated to ensure the final PDFs are statistically robust. Ultimately, 100 random redshift samples are drawn from the GMM of each of the 100 iterations. The resulting 10,000 samples are converted into a quantile-based PDF and compressed using the \textsc{ColdPress} module.

\section{Results}
\label{sec:results}

\subsection{Best photo-z point estimate}
\label{sec:best-point}

\begin{figure}
\begin{center}
\includegraphics[width=\columnwidth]{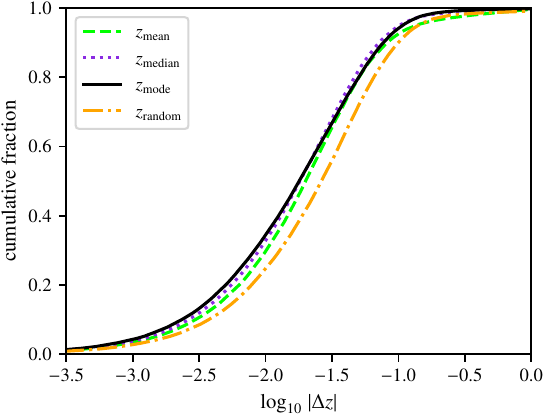}
\end{center}
\vspace{-0.4cm}
\caption[]{Cumulative fraction of sources with redshift error $|\Delta z|$ = |\zphot-\zspec|/(1+\zspec) smaller than a given threshold for different point estimates. The sample consists of galaxies in the \texttt{jplus+wise} test set with \zspec\ $<$ 1. \label{fig:logdz-distrib-glx1}}
\end{figure}

\begin{figure}
\begin{center}
\includegraphics[width=\columnwidth]{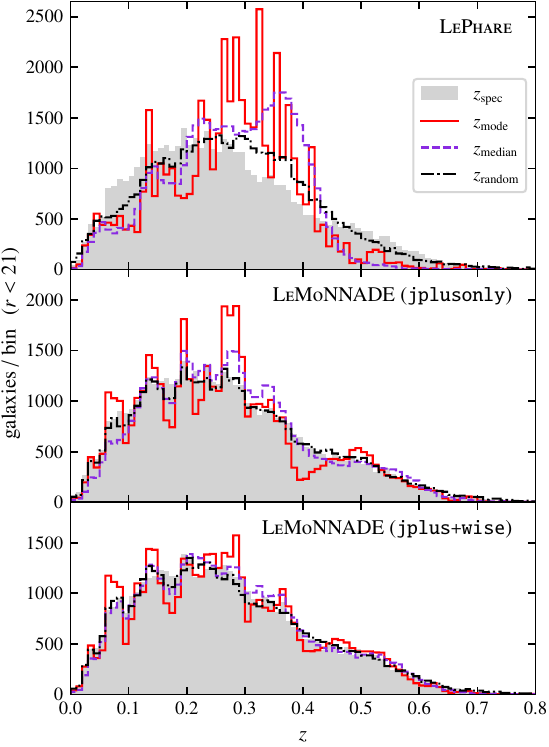}
\end{center}
\vspace{-0.4cm}
\caption[]{Redshift distributions derived from different PDF point estimates compared to the spectroscopic distribution (grey shaded area) for $r$$<$21 galaxies in the test sample.\label{fig:zdistrib-glx1}}
\end{figure}

\begin{figure*}
\begin{center}
\includegraphics[width=\textwidth]{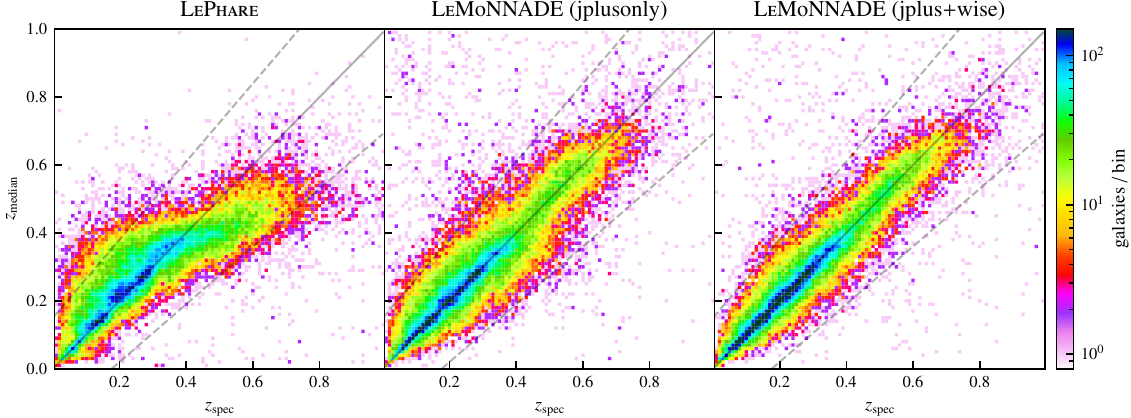}
\end{center}
\vspace{-0.4cm}
\caption[]{Photometric vs spectroscopic redshift for $z<1$ galaxies in the test sample. The colour coding indicates the logarithmic density of galaxies in bins of $0.01 \times 0.01$. The solid diagonal marks the 1:1 relation, while dashed lines indicate the catastrophic outlier threshold $|\Delta z| = 0.15$. \label{fig:zphot-zspec-map}}
\end{figure*}

While the full PDF contains the complete information regarding the redshift uncertainty of a source, scalar point estimates are often required for specific scientific applications. \textsc{ColdPress} derives several such estimates from the compressed PDFs: the mode (\zmode), median (\zmedian), mean (\zmean), and a Monte Carlo sample drawn randomly from the cumulative distribution (\zrandom).

The choice of the optimal point estimate depends on the intended application. As shown in Fig.~\ref{fig:logdz-distrib-glx1}, \zmode\ maximises the probability of obtaining a small redshift error ($|\Delta z|$) while \zmedian\ is slightly more effective at suppresing catastrophic outliers. Both \zmean\ and \zrandom\ exhibit significantly lower precision in terms of $|\Delta z|$ and are therefore less suitable for applications requiring high accuracy for individual objects. 

Figure~\ref{fig:zdistrib-glx1} highlights a critical systematic affecting the distribution of \zmode: strong ``aliasing'' artefacts, characterised by artificial overdensities at specific redshifts at the expense of adjacent intervals. This aliasing \citep[also known as ``redshift focusing'', e.g.][]{Wolf2009, Hildebrandt2012} is a consequence of sampling a continuous spectrum with a discrete set of photometric bands. 
The photo-$z$ likelihood function has local maxima at redshifts where noise-induced spikes in the photometric SED align with real spectral features. While this phenomenon applies to all filtersets and photometric redshift methods, it is particularly strong in J-PLUS due to its specific design featuring a few narrow bands separated by significant spectral gaps. We also observe that aliasing is more severe in the \textsc{LePhare} results compared to \textsc{LeMoNNADE}, likely because the template-fitting approach builds a coarser mapping of the colour-redshift space using a limited set of fixed templates.

The \zmode\ estimator is more sensitive to these noise-driven fluctuations than \zmedian, as the peak position of a PDF is less stable than its integrated median value. Consequently, \zmedian\ provides a redshift distribution that is significantly smoother and less biased by aliasing artefacts. 
While \zrandom\ is the least accurate point estimate for individual objects, it provides the only estimator that statistically reproduces the full PDF of the population. Accordingly, \zrandom\ is also the point estimate least affected by aliasing effects. For the sake of brevity, the subsequent evaluation of photometric redshift accuracy for individual sources relies exclusively on \zmedian. We note, however, that all identified trends apply equally to \zmode. Finally, in Sect.~\ref{sec:z-distribution}, we use \zrandom\ to derive the unbiased redshift distribution, $N(z)$, for the entire J-PLUS sample.

\subsection{Photo-z accuracy for galaxies}

\begin{figure}
\begin{center}
\includegraphics[width=\columnwidth]{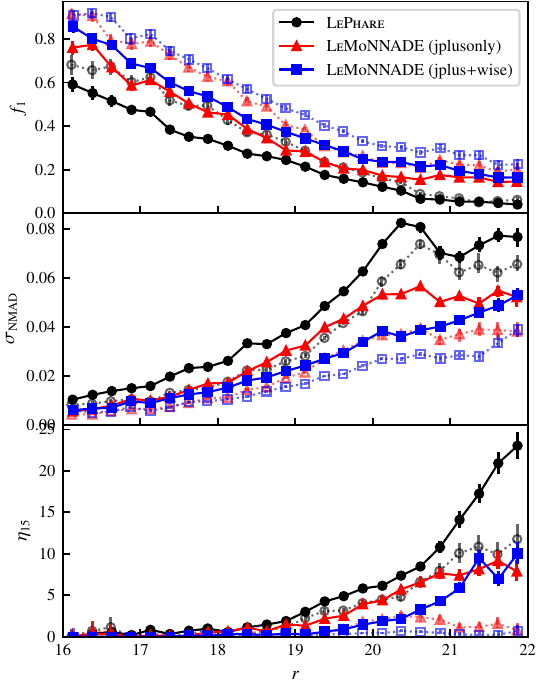}
\end{center}
\vspace{-0.4cm}
\caption[]{Magnitude dependence of the $f_1$ metric (top), \snmad\ (middle), and $\eta_{15}$ (bottom) for \textsc{LePhare} and the two \textsc{LeMoNNADE} configurations. Solid symbols represent the full sample; open symbols correspond to the 50\% of sources with the highest \texttt{odds} in each magnitude bin. \label{fig:metrics-glx1}}
\end{figure}

\begin{figure*}
\begin{center}
\includegraphics[width=\textwidth]{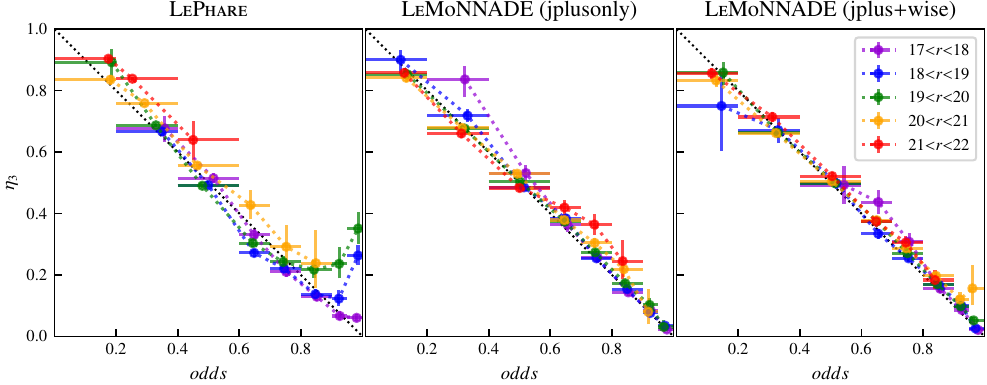}
\end{center}
\vspace{-0.4cm}
\caption[]{Observed outlier rate as a function of the mean \texttt{odds} parameter. The dotted line represents the theoretical expectation for perfectly calibrated probabilities. \label{fig:outrate-by-odds}}
\end{figure*}

We compare the performance of the template-fitting code \textsc{LePhare} against the machine-learning pipeline \textsc{LeMoNNADE}. To ensure a fair comparison, we restrict the evaluation set to the subset of the test sample with spectroscopic classification as ``galaxy'' and \zspec\ $<$ 1, matching the redshift search space imposed on the \textsc{LePhare} configuration (see Sect.~\ref{sec:lephare}). This subset contains 50,325 sources.

Figure~\ref{fig:zphot-zspec-map} displays the density of galaxies in the photometric versus spectroscopic redshift plane. A logarithmic scale is used to highlight the outliers. The \textsc{LeMoNNADE} predictions follow the 1:1 relation tightly across the entire redshift range. In contrast, the \textsc{LePhare} photo-$z$ estimates show a bias towards \zmedian\ $>$ \zspec\ at low redshift and \zmedian\ $<$ \zspec\ at high redshift. This behaviour is likely a consequence of the redshift prior in \textsc{LePhare}, which favours intermediate values ($z \sim 0.4$--$0.5$) for sources at faint magnitudes. 

The dependence of photo-$z$ quality on magnitude is quantified in Fig.~\ref{fig:metrics-glx1}, which presents the precision metric \fone\ (the fraction of sources with redshift errors $|\Delta z| < 0.01$), the robust scatter \snmad, and the catastrophic outlier rate \etafifteen\ ($|\Delta z| > 0.15$). We show results considering all the sources in each magnitude bin (solid symbols) as well as results obtained after applying a quality cut that takes, from each magnitude bin, the top half of the \texttt{odds} distribution (open symbols).

\textsc{LeMoNNADE} consistently outperforms \textsc{LePhare} across all metrics and magnitude ranges, with the performance gap widening significantly at the faint end. This indicates that the neural network captures the colour-redshift mapping more effectively than the discrete template set used in \textsc{LePhare}, likely due to its ability to interpolate across the parameter space and exploit subtle features in the low-S/N regime.

Comparing the two \textsc{LeMoNNADE} configurations, the inclusion of WISE photometry (\texttt{jplus+wise}) yields a significant reduction in the catastrophic outlier rate at intermediate magnitudes ($18 < r < 21$) compared to the optical-only dataset (\texttt{jplusonly}). This improvement also translates into a lower \snmad\ and higher \fone. At the faint limit ($r \sim 22$), the metrics for both datasets converge, as the majority of galaxies in this range fall below the detection limit of the unWISE catalogue.

We also assess the reliability of the probabilistic outputs by comparing the outlier rate \etathree\ (the fraction of sources with redshift errors $|\Delta z| > 0.03$) with the average value of the \texttt{odds} parameter in bins of \texttt{odds} (Fig.~\ref{fig:outrate-by-odds}). For realistic PDFs, the expectation is $\eta_{3} \approx 1 - \langle\rm{odds}\rangle$. The \texttt{odds} parameter from \textsc{LePhare} is generally well-calibrated, although it tends towards overconfidence at high values (\texttt{odds} $\gtrsim 0.9$). Conversely, \textsc{LeMoNNADE} produces highly realistic probabilities at the high-confidence end but appears slightly overconfident at intermediate \texttt{odds} values.

\subsection{Photo-z accuracy for QSOs}

We evaluate the photometric redshift accuracy of {\sc LeMoNNADE} for quasars using the 7,774 sources in the test sample with a spectroscopic classification of `QSO'.
The inclusion of infrared photometry from WISE (the \texttt{jplus+wise} dataset) proves to be strongly beneficial for this class of objects.
Specifically, the rate of catastrophic outliers, \etafifteen, is reduced from $\sim$40\% in the \texttt{jplusonly} configuration to $\sim$23\% for \texttt{jplus+wise}.
Consequently, we focus the subsequent analysis exclusively on the results obtained with the \texttt{jplus+wise} configuration.

\begin{figure}
\begin{center}
\includegraphics[width=\columnwidth]{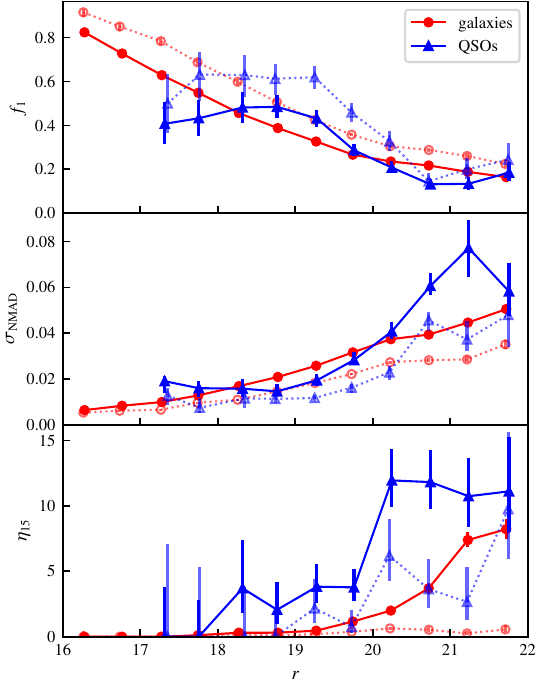}
\end{center}
\vspace{-0.4cm}
\caption[]{Magnitude dependence of the $f_1$ metric (top), \snmad\ (middle), and $\eta_{15}$ (bottom) for the photometric redshifts of QSOs (triangles) and galaxies (circles) from the spectroscopic test sample, obtained with \textsc{LeMoNNADE} in the \texttt{jplus+wise} configuration.
Solid symbols represent the full sample; open symbols correspond to the 50\% of sources with the highest \texttt{odds} in each magnitude bin. \label{fig:metrics-glx-qso}}
\end{figure}

Figure~\ref{fig:metrics-glx-qso} presents the accuracy metrics for QSOs compared to galaxies as a function of magnitude.
For bright sources ($r < 20$), the precision ($f_1$) and robust scatter (\snmad) of QSOs are comparable to those of galaxies at similar magnitudes.
However, QSOs have a significantly higher fraction of catastrophic outliers.
This disparity is largely attributable to the wide redshift range occupied by the quasar population (extending up to $z \sim 4$), whereas 99\% of the galaxies with $r < 20$ are located at $z < 0.5$.

\begin{figure}
\begin{center}
\includegraphics[width=\columnwidth]{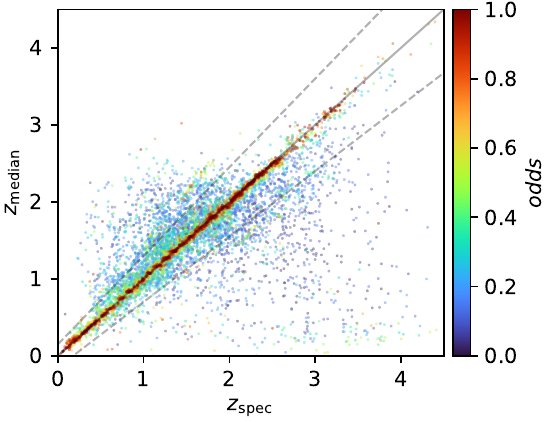}
\end{center}
\vspace{-0.4cm}
\caption[]{Photometric versus spectroscopic redshift for QSOs in the test sample.
The colour coding indicates the \texttt{odds} parameter corresponding to the point estimate $z_{\rm{median}}$.
The solid diagonal marks the 1:1 relation, while dashed lines indicate the catastrophic outlier threshold $|\Delta z| = 0.15$. \label{fig:zphot-zspec-qsos}}
\end{figure}

As shown in Fig.~\ref{fig:zphot-zspec-qsos}, most of these QSO outliers have very low values of the \texttt{odds} parameter, which is indicative of broad PDFs likely as a consequence of the absence of strong emission lines at the wavelengths probed by the narrow-band filters of J-PLUS. In the near future, the combination of J-PLUS with the near-IR spectrophotometric data from the SPHEREx survey
\citep{Bock2026} will significantly improve the photo-z accuracy and
reduce the fraction of catastrophic outliers for QSOs, providing an
unprecedented dataset with both a broad wavelength coverage and low-
resolution spectral information.

\subsection{Redshift bias}

\begin{figure}
\begin{center}
\includegraphics[width=\columnwidth]{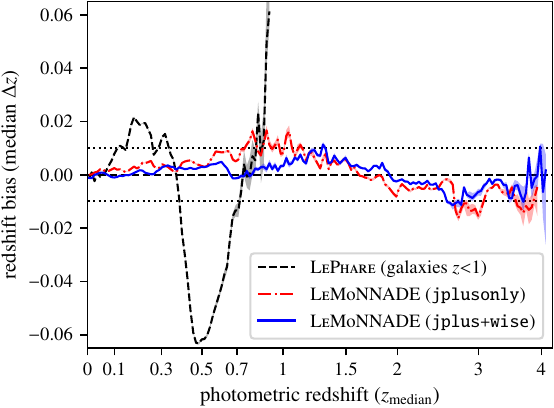}
\end{center}
\vspace{-0.4cm}
\caption[]{Redshift dependence of the photometric redshift bias, measured in bins of width $0.01(1+z)$. Broken lines represent the median $\Delta z$ within each bin, excluding catastrophic outliers, while shaded regions denote the corresponding 1-$\sigma$ confidence intervals.\label{fig:photoz-bias}}
\end{figure}

For specific scientific applications, systematic biases in photo-$z$ are more determinant than the precision achieved for individual sources. We define the photometric redshift bias as the median of the redshift error, $\Delta z = (z_{\rm{phot}} - z_{\rm{spec}})/(1+z_{\rm{spec}})$, computed in bins of $z_{\rm{phot}}$ after excluding catastrophic outliers ($|\Delta z| > 0.15$). The extensive spectroscopic sample enables the measurement of this bias with high redshift resolution and low statistical uncertainty. 

Figure~\ref{fig:photoz-bias} shows the median $\Delta z$ and its 1-$\sigma$ uncertainty as a function of photometric redshift. The \textsc{LePhare} estimates, calculated exclusively for galaxies at $z<1$, have substantial systematic offsets reaching up to 6\%. In contrast, the bias of \textsc{LeMoNNADE} photo-z in the same range is $<$2\% for the \texttt{jplusonly} configuration and $<$0.5\% for \texttt{jplus+wise}. The latter also stays within $\pm$1\% across the entire redshift range up to $z=4$.

\subsection{Binary classification}
\label{sec:accuracy-binclass}

\begin{figure}
\begin{center}
\includegraphics[width=\columnwidth]{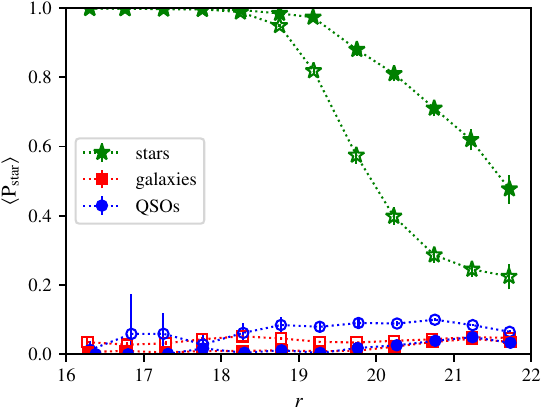}
\end{center}
\vspace{-0.4cm}
\caption[]{Mean \pstar{} as a function of $r$-band magnitude for stars (green), galaxies (red), and QSOs (blue). Open and solid symbols denote the \texttt{jplusonly} and \texttt{jplus+wise} datasets, respectively. \label{fig:pstar-distrib}}
\end{figure}

\begin{figure}
\begin{center}
\includegraphics[width=\columnwidth]{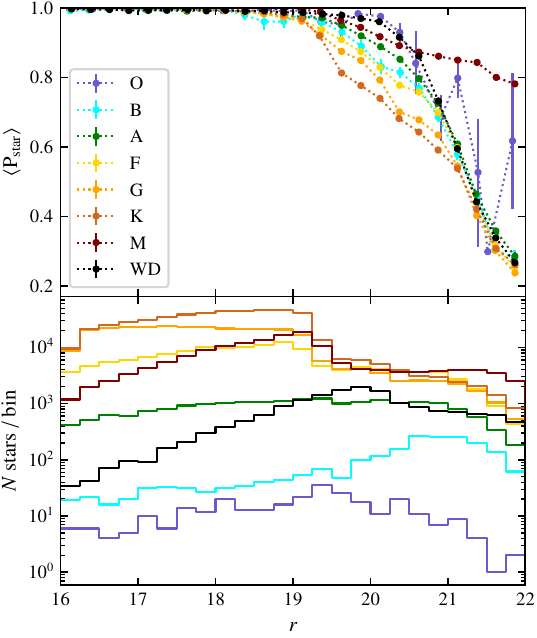}
\end{center}
\vspace{-0.4cm}
\caption[]{Classification uncertainty and sample abundance by spectral type.
Top panel: Mean probability of being a star, $\langle P_{\rm{star}} \rangle$, as a function of $r$-band magnitude for spectroscopically confirmed stars, differentiated by spectral class.
Bottom panel: Number of sources per magnitude bin for each spectral class in the spectroscopic training sample.
\label{fig:startype}}
\end{figure}

\begin{figure}
\begin{center}
\includegraphics[width=\columnwidth]{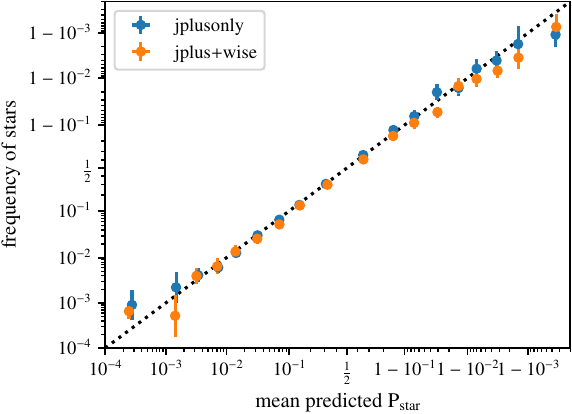}
\end{center}
\vspace{-0.4cm}
\caption[]{Calibration test for \pstar\ values: fraction of spectroscopic stars versus predicted \pstar. \label{fig:pstar-calibration}}
\end{figure}

We evaluate the discriminative power of our {\sc LeMoNNADE} classification by obtaining average \pstar\ values as a function of magnitude for spectroscopically confirmed stars, galaxies and QSOs (Figure~\ref{fig:pstar-distrib}).
The inclusion of WISE photometry significantly enhances discrimination, specially for stars and QSOs, due to colour degeneracy in the optical bands.
The classification uncertainty for stars shows a stronger dependence on magnitude compared to QSOs, likely driven by the class imbalance in the training sample, which is dominated by stars at bright magnitudes and by QSOs at the faint end (Fig. \ref{fig:mag-distrib}).

We further investigate the classification uncertainty by separating the sample into spectral types (Fig.~\ref{fig:startype}.)
While M-dwarfs remain robustly identified with $\langle P_{\rm{star}} \rangle \gtrsim$ 0.8 across the entire magnitude range (likely due to their strong molecular absorption features), the remaining spectral types show a marked decline in \pstar\ at faint magnitudes.
Interestingly, in the intermediate magnitude regime ($19 \lesssim r \lesssim 21$), we observe a clear correlation between effective temperature and \pstar, where hotter stars (types O, B, and WD) achieve higher average \pstar\ than cooler stars (types G and K).
This trend runs contrary to the demographics of the training sample, where G and K stars are orders of magnitude more abundant than early-type stars (Fig.~\ref{fig:startype}, bottom panel).
This suggests that degeneracy with low-redshift galaxies in the colour space, exacerbated by the lower $S/N$ in the blue J-PLUS filters, drives the higher classification uncertainty for faint G and K stars.

We test the calibration of \pstar\ values by computing the frequency of spectroscopic stars as a function of the probability \pstar\ (Fig.~\ref{fig:pstar-calibration}). Classifications are unbiased across the entire dynamic range for both the \texttt{jplusonly} and \texttt{jplus+wise} datasets, except for some overconfidence at the extreme ends of the distribution (\pstar\ $< 0.001$ or $> 0.999$); however, this could also be caused by impurity of the spectroscopic training labels at a level of $\sim$0.1\%. 

\section{Properties of the J-PLUS extragalactic population}
\label{sec:jplus-population}

Having evaluated the model predictions against the spectroscopic control sample, we now characterise the statistical properties of the full J-PLUS extragalactic population using {\sc LeMoNNADE} with the \texttt{jplus+wise} configuration. Applying models trained on spectroscopic surveys to a magnitude-limited sample introduces systematic challenges, as the predictions carry the selection biases of the training data. Specifically, target selection criteria from SDSS and DESI strongly bias against stars, particularly at the faint end. Applying these raw priors directly would severely underestimate faint stellar counts. We rectify this by calibrating the raw \pstar\ probabilities using an Expectation-Maximization algorithm, as detailed in Appendix \ref{sec:pstar-calibration}. Furthermore, analysing the full photometric sample requires accounting for non-astrophysical contaminants. In Appendix \ref{sec:remove-fakes}, que quantify the probability of being a false detection (\pnoise) for individual sources based on the S/N of their forced photometry.
Given the considerable size of the full catalogue, we perform the subsequent analysis on a randomly selected subsample comprising 10\% of the data\footnote{Specifically, we select sources where the \texttt{NUMBER} identifier ends in the digit 9. Since \texttt{SExtractor} assigns these identifiers sequentially based on the Y-coordinate of detections within each tile, this modulo-based selection ensures a spatially uniform and statistically representative subset of the survey population.}.

\subsection{Number density}
\label{sec:number-density}

\begin{figure}
\begin{center}
\includegraphics[width=\columnwidth]{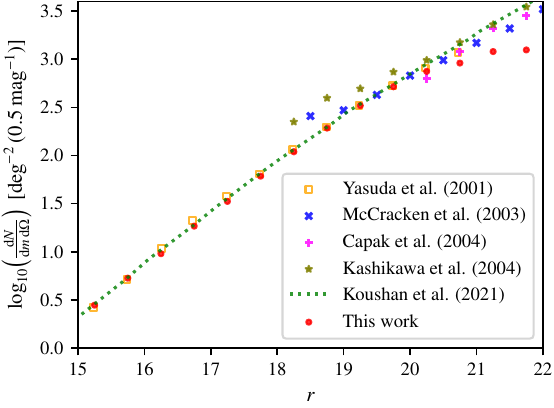}
\end{center}
\vspace{-0.4cm}
\caption[]{Differential number counts of J-PLUS extragalactic sources as a function of extinction-corrected $r$-band magnitude. 
For comparison, we include galaxy counts from SDSS \citep{Yasuda2001} and \citet{Koushan2021}. 
Additional reference counts are derived from \citet{McCracken2003} (CFH12K-VIRMOS deep field), \citet{Kashikawa2004} (Subaru Deep Field), and \citet{Capak2004} (Hawaii HDF-N).
The reference measurements were originally taken in the $R$ band and subsequently converted to the $r$ band by N.~Metcalfe (available at \url{https://astro.dur.ac.uk/~nm/pubhtml/counts/counts.html}).
\label{fig:counts-galaxies}}
\end{figure}

In Figure~\ref{fig:counts-galaxies}, we show the extragalactic counts in J-PLUS as a function of the extinction-corrected $r$-band magnitude, computed in bins of 0.5~mag to facilitate comparison with literature data. 
To minimise the impact of stellar misclassification in extragalactic counts, we exclude J-PLUS tiles with Galactic latitudes $|b| < 15^\circ$, as well as a few tiles with unusually shallow observations and those covering the disc of M31, which have a higher rate of spurious detections (see Appendix~\ref{sec:highz-excess}). 
The excluded tiles represent 20\% of the total J-PLUS DR4 footprint. 

The extinction-corrected J-PLUS extragalactic counts show excellent agreement with the SDSS galaxy counts from \citet{Yasuda2001}, except in the $16 < r < 18$ interval where SDSS counts are slightly higher. 
This discrepancy is noteworthy given that the J-PLUS counts include quasars, which are expected to contribute approximately 3--5\% of the extragalactic sources in this magnitude range \citep{Richards2006}. 
An independent determination of galaxy number counts by \citet{Koushan2021}, using a combination of datasets from GAMA, DEVILS, and several HST surveys, suggests that the SDSS counts could be overestimated in this regime. 
At fainter magnitudes ($r \gtrsim 20.5$--$21.0$), the J-PLUS extragalactic counts flatten, likely reflecting a decrease in detection completeness that is more pronounced for extended sources. 
Compared to the relation from \citet{Koushan2021}, J-PLUS counts are lower by $\sim$10\% at $r=20$ and $\sim$40\% at $r=21$.

\subsection{Redshift distribution}
\label{sec:z-distribution}

Under the assumption that the redshift PDFs are well calibrated, the true redshift distribution, $N(z)$, of the extragalactic J-PLUS population is theoretically the sum of the individual PDFs. Since evaluating millions of full PDFs is computationally expensive, we estimate $N(z)$ using a Monte Carlo approach by aggregating a single random draw, \zrandom, from the PDF of each source. Given the immense sample size, the distribution of \zrandom\ statistically converges to the exact sum of the PDFs. The differential number density per unit solid angle in a redshift bin of width $\Delta z$ is computed as:

\begin{equation}
\label{eq:dN-raw}
\frac{dN}{dz\,d\Omega} \approx \frac{1}{A_{\rm eff}\Delta z} \sum_{i} w_i \, \mathcal{I}(z \le z_{{\rm random}, i} < z + \Delta z) 
\end{equation}

where $A_{\rm eff}$ represents the effective survey area, $w_i = 1 - P_{\rm star,i}$ is the probability that source $i$ is extragalactic (derived from its calibrated \pstar\ value), the index $i$ extends over all J-PLUS sources with $r < 21$, and $\mathcal{I}$ is an indicator function that equals 1 if the condition is met and 0 otherwise.

\begin{figure}
\begin{center}
\includegraphics[width=\columnwidth]{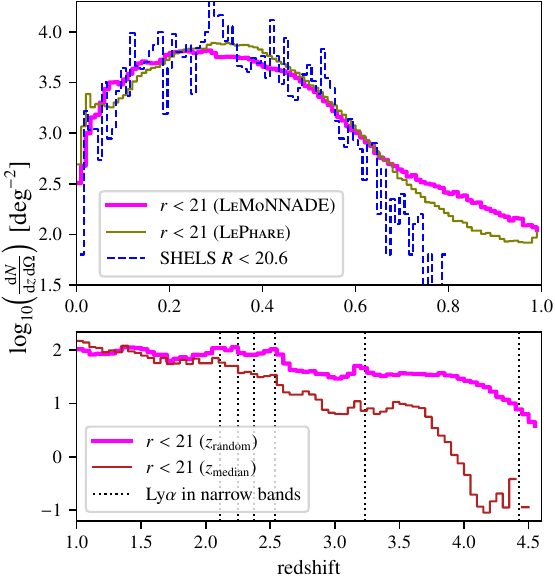}
\end{center}
\vspace{-0.4cm}
\caption[]{Differential number counts as a function of redshift for the full $r < 21$ J-PLUS sample, obtained by weighting each source by its probability of being extragalactic. Top panel: The $z < 1$ distributions obtained from the full PDFs (\zrandom) from \textsc{LePhare} (thin solid line) and \textsc{LeMoNNADE} (using the \texttt{jplus+wise} dataset, thick solid line). For comparison, counts from the spectroscopic SHELS survey are also shown. Bottom panel: Differential number counts at $z > 1$ obtained with \textsc{LeMoNNADE} (\texttt{jplus+wise} dataset) using either \zrandom\ or the median of the PDF (\zmedian). Vertical dotted lines denote the redshifts at which the Ly$\alpha$ emission line falls within the transmission window of the narrow J-PLUS filters.\label{fig:Nz-comparison}}
\end{figure}

 The resulting redshift distribution derived from the photo-$z$ of \textsc{LeMoNNADE} with the \texttt{jplus+wise} dataset is shown as the thick magenta line in Fig.~\ref{fig:Nz-comparison}. The number density presents a broad peak at $z \sim 0.3$, with 50\% of the population contained within the interval $0.1 < z < 0.57$. The high-redshift tail, populated primarily by quasars, extends to $z > 4$ (Fig.~\ref{fig:Nz-comparison}, bottom panel).

The redshift distribution obtained with \textsc{LePhare}, which assumes all extragalactic sources are galaxies at $z < 1$, broadly matches the \textsc{LeMoNNADE} results, but peaks at a slightly higher redshift ($z \sim 0.35$) and has lower counts at $z \gtrsim 0.7$. For comparison, we show the redshift distribution from the Smithsonian Hectospec Lensing Survey \citep[SHELS;][]{Geller2014}, which covers an effective area of 3.98 deg$^2$ and is 95\% complete to $R = 20.6$. Although the SHELS distribution is noisy due to the significant cosmic variance expected for such a small effective area, its general trend is consistent with both \textsc{LePhare} and \textsc{LeMoNNADE} estimates up to $z \sim 0.6$. Beyond $z = 0.6$, a growing deficit in SHELS galaxy counts relative to J-PLUS emerges. This discrepancy may partially arise from the slightly shallower magnitude limit of the SHELS survey ($R = 20.6$ compared to $r = 21.0$) and its 5\% incompleteness, factors that disproportionately affect faint galaxies at higher redshifts. Alternatively, the J-PLUS counts at $z > 0.6$ could be systematically overestimated by both \textsc{LePhare} and \textsc{LeMoNNADE}. We investigate this possibility in Appendix \ref{sec:highz-excess}, demonstrating that contamination from spurious sources contributes significantly to the observed excess in J-PLUS counts at $z > 0.6$.

The bottom panel of Fig.~\ref{fig:Nz-comparison} shows the redshift distribution from \textsc{LeMoNNADE} at $z > 1$. At $r < 21$, this redshift regime is dominated by QSOs, with stars and scattered lower-redshift galaxies as potential contaminants. Unlike the low-redshift behaviour, the number counts at $z \gtrsim 2$ estimated using the full PDF (\zrandom) exceed those obtained using the median redshift (\zmedian). This discrepancy is not explained by uncertainty in the photo-$z$ of QSOs (\snmad\ $< 0.06$, \etafifteen\ $< 10\%$). Instead, the excess in $z_{\mathrm{random}}$ counts is driven by the low-probability, high-redshift tails in the PDFs of the $z < 1$ galaxy population. Because these low-redshift galaxies are approximately 20 times more abundant than QSOs, even a small probability density extending to high redshift introduces a substantial number of artificial counts when the full PDF is stochastically sampled. Consequently, $z_{\mathrm{median}}$ effectively suppresses this low-probability noise, providing a more accurate representation of the true $N(z)$ for the high-redshift QSO population.

Both \zrandom\ and \zmedian\ show overdensities at $z > 2$ that coincide with the redshifts at which the Ly$\alpha$ emission line shifts into the transmission windows of one of the J-PLUS narrow-band filters. This clustering is driven by a systematic variation with redshift of the photo-$z$ accuracy. Because QSOs typically have a power-law-like continuum, their redshift determination is highly dependent on detecting a few prominent emission lines. When Ly$\alpha$ falls within a narrow filter, the photometric constraint is strong, yielding sharply peaked, narrow PDFs. Conversely, when the line falls outside these filters, the lack of distinct features results in broader PDFs. This modulation in PDF width naturally focuses the point estimates, creating apparent overdensities at the highly-constrained redshifts.

Under the assumption that the population with \zmedian\ $> 1.2$ is dominated by quasars, we estimate a surface density of $\sim$100 deg$^{-2}$ for QSOs with $r < 21$. This value is nearly double the density of $\sim$55 deg$^{-2}$ reported by \citet{Wolf2003} for QSOs in the COMBO-17 survey with $z_{\rm phot} > 1.2$ and $R < 21$. However, we caution that our estimate may be biased by residual contamination from faint blue stars in the Galactic halo. While we demonstrate in Fig.~\ref{fig:zdistrib-jpsv} that stellar contamination does not significantly affect the galaxy redshift distribution, its impact on the quasar population remains difficult to quantify, as stars and quasars are morphologically indistinguishable. Notably, applying a stricter purity threshold of \pstar\ $< 0.1$ reduces the estimated density to $\sim$53 deg$^{-2}$, a value fully consistent with the COMBO-17 estimate.

We define high-confidence extragalactic samples by selecting sources with $r < 21$ and a calibrated stellar probability \pstar\ $< 0.1$. From this population, we extract a galaxy-dominated sample (\zmedian\ $< 0.6$) and a QSO-dominated sample (\zmedian\ $> 1.2$). 
The typical photometric redshift uncertainty for these sources is quantified by the robust scatter, \snmad$(\Delta z)$, where $\Delta z = (z_{\rm median} - z_{\rm spec})/(1 + z_{\rm spec})$. Because the majority of sources in these samples lack spectroscopic redshifts, we estimate \snmad$(\Delta z)$ by substituting \zspec\ with \zrandom. This substitution is theoretically robust: assuming the PDFs are well-calibrated, \zrandom\ correctly samples the true underlying redshift probability space. Consequently, the distribution of $(z_{\rm median} - z_{\rm random})/(1 + z_{\rm random})$ is statistically equivalent to the distribution of the true error, $(z_{\rm median} - z_{\rm spec})/(1 + z_{\rm spec})$.
Using this estimation, the high-confidence galaxy sample comprises $8.34 \times 10^6$ sources with \snmad$(\Delta z) \sim 0.032$. The high-confidence QSO sample contains $2.30 \times 10^5$ sources with \snmad$(\Delta z) \sim 0.043$.

\section{Conclusions}
\label{sec:conclusions}

In this work, we have presented the value-added catalogues of spectral classification and photometric redshifts for Data Release 4 (DR4) of the J-PLUS survey. 
This dataset covers an effective area of 4437.5 deg$^2$ and comprises 79.2 million unique sources up to a magnitude limit of $r < 22$. 
Leveraging the 12-band photometry of J-PLUS, which combines broad and narrow filters, we have generated redshift estimates and class probabilities using two complementary methodologies: a template-fitting approach using {\sc LePhare} and a novel machine learning pipeline named {\sc LeMoNNADE}. 

We found that the machine learning approach, {\sc LeMoNNADE}, consistently outperforms the template-fitting method in terms of precision (\fone), robust scatter (\snmad), and outlier rate (\etafifteen) across all magnitude ranges. 
    This performance gap widens at the faint end, suggesting that the neural network, aided by spectral mixing augmentation, more effectively captures the complex mapping between noisy photometry and redshift than discrete template sets. 

We provide a probabilistic star/galaxy separation that is unbiased across the full dynamic range. 
    We addressed the inherent selection bias of the spectroscopic training samples (which under-represent stars) by implementing a Bayesian calibration based on an Expectation-Maximization algorithm. 
    The resulting classifications show that intrinsic spectral energy distribution features and signal-to-noise ratio, rather than training set composition, are the dominant factors driving classification accuracy at the faint end. 
The inclusion of infrared photometry from WISE (W1 and W2 bands) significantly enhances the accuracy of {\sc LeMoNNADE} photo-z and spectral classification.

We have characterised the J-PLUS extragalactic population. Source counts as a function of the extinction-corrected $r$ magnitude show excellent agreement with literature results, but flatten at $r \gtrsim 20.5$ due to detection incompleteness.
The inferred redshift distribution for the extragalactic population peaks at $z \sim 0.3$, with a high-redshift tail dominated by QSOs extending to $z \sim 4$. 
The distribution shows apparent over-densities at redshifts where the Lyman-$\alpha$ emission line shifts into the J-PLUS narrow-band filters. We attribute this effect to the variation in photo-$z$ precision: sources with strong spectral features falling within narrow bands have better-constrained PDFs, resulting in apparent clustering in the redshift distribution. 
 
Finally, we isolate high-purity extragalactic samples (\pstar\ $< 0.1$) comprising $8.34 \times 10^6$ galaxies and $2.30 \times 10^5$ quasars. Their estimated typical redshift errors are \snmad\ = 0.032 and 0.043, respectively.    

\section*{Data availability}
The J-PLUS DR4 spectral classifications and photometric redshifts presented in this article are publicly available via the CEFCA Catalogues portal (\url{https://archive.cefca.es/catalogues}) and through its Virtual Observatory (VO) Table Access Protocol (TAP) service at \url{https://archive.cefca.es/catalogues/vo/tap/jplus-dr4}. The data can be queried from the tables \texttt{jplus.PhotoZLephare}, \texttt{jplus.PhotoZLemonnade}, and \texttt{jplus.PhotoZLemonnadeWise}. The \textsc{ColdPress} software (v1.1.2) used in this work is publicly available on GitHub at \url{https://github.com/ahc-photoz/coldpress}.

\begin{acknowledgements}

We thank A.~Alvarez-Candal and P.~T.~Rahna for comments during the internal collaboration review.

Financial support from the Spanish Ministerio de Ciencia, Innovaci\'on y Universidades (MCIU/AEI/10.13039/501100011033) is acknowledged through the following grants: PID2023-147386NB-I00 (A.H.-C., co-funded by ERDF/EU); CNS2023-145339 (J.A.F.O., co-funded by EU NextGenerationEU/PRTR); RyC2022-030469-I (H.D.S.); CEX2021-001131-S (\emph{Severo Ochoa}), PID2024-155572NB-C22 (co-funded by FEDER/EU), CNS2025-166660, and RYC2022-038448-I (co-funded by the European Social Fund Plus) (A.dP); and PID2023-153342NBI00 (J.A.L.A., co-funded by ERDF). J.A.F.O. additionally acknowledges support from MCIN and the EU NextGenerationEU through project ICTS-MRR-2021-03-CEFCA. 
A.dP acknowledges the RyC-MAX grant 20245MAX008 (CSIC). V.M. thanks CNPq (Brazil), FAPES (Brazil), and CAPES (Brazil) for partial financial support. 
Based on observations made with the JAST80 telescope and T80Cam camera for the J-PLUS project at the Observatorio Astrof\'{\i}sico de Javalambre (OAJ), in Teruel, owned, managed, and operated by the Centro de Estudios de F\'{\i}sica del  Cosmos de Arag\'on (CEFCA). We acknowledge the OAJ Data Processing and Archiving Unit (UPAD; \citealt{upad}) for reducing the OAJ data used in this work.
Funding for the J-PLUS Project has been provided by the Governments of Spain and Arag\'on through the Fondo de Inversiones de Teruel; the Aragonese Government through the Research Groups E96, E103, E16\_17R, E16\_20R, and E16\_23R;  grants PID2024-162229NB-I00, PID2024-155572NB-C21, and PID2024-155572NB-C22 (MICIU/AEI/10.13039/501100011033 and ERDF/EU); the Spanish Ministry of Science and Innovation (MCIN/AEI/10.13039/501100011033 y FEDER, Una manera de hacer Europa) with grants PID2021-124918NB-C41, PID2021-124918NB-C42, PID2021-124918NA-C43, and PID2021-124918NB-C44; the Spanish Ministry of Science, Innovation and Universities (MCIU/AEI/FEDER, UE) with grants PGC2018-097585-B-C21 and PGC2018-097585-B-C22; the Spanish Ministry of Economy and Competitiveness (MINECO) under AYA2015-66211-C2-1-P, AYA2015-66211-C2-2, AYA2012-30789, and ICTS-2009-14; and European FEDER funding (FCDD10-4E-867, FCDD13-4E-2685). The Brazilian agencies FINEP, FAPESP, and the National Observatory of Brazil have also contributed to this project.
The authors gratefully acknowledge the computer resources at Artemisa and the technical support provided by the Instituto de Fisica Corpuscular, IFIC(CSIC-UV). Artemisa is co-funded by the European Union through the 2014-2020 ERDF Operative Programme of Comunitat Valenciana, project IDIFEDER/2018/048.

\end{acknowledgements}

\appendix

\section{Calibration of \pstar\ values}
\label{sec:pstar-calibration}

The \pstar\ values generated by {\sc LeMoNNADE} estimate the probability that a source is a star, conditioned on the feature space of the training data. However, these estimates implicitly incorporate the class priors of the spectroscopic training set ($\pi_{\rm train}$), which suffers from severe selection effects. Specifically, the spectroscopic target selection of surveys such as SDSS and DESI introduces a strong bias against stars relative to their true abundance in a magnitude-limited photometric survey ($\pi_{\rm full}$). Consequently, the raw \pstar\ values are generally underestimated for the full J-PLUS population, particularly at faint magnitudes. In this regime, the training sample is not only dominated by extragalactic sources, but the discrimination between stars and galaxies is also hindered by larger photometric uncertainties. As the observed spectral features become less distinct due to noise, the predictions of the model rely more heavily on the learned (biased) priors, thereby exacerbating the discrepancy.

To recover the posterior probabilities applicable to the unbiased J-PLUS source population, we apply a Bayesian correction that shifts the probabilities from the training prior to the target prior. The calibrated probability, $p_{\rm cal}$, is related to the raw probability from {\sc LeMoNNADE}, $p_{\rm raw}$, via:

\begin{equation}
\label{eq:adjust-pstar-prob}
p_{\rm cal} = \frac{p_{\rm raw} \frac{\pi_{\rm full}}{\pi_{\rm train}} }{ p_{\rm raw} \frac{\pi_{\rm full}}{\pi_{\rm train}} + (1 - p_{\rm raw}) \frac{1 - \pi_{\rm full}}{1 - \pi_{\rm train}} }
\end{equation}

\noindent where $\pi_{\rm train}$ is the frequency of stars in the spectroscopic training sample and $\pi_{\rm full}$ is the (unknown) frequency of stars in the full J-PLUS catalogue.

We estimate $\pi_{\rm full}$ and derive $p_{\rm cal}$ simultaneously using the Expectation-Maximization algorithm described by \citet{Saerens2002}. This iterative procedure maximises the likelihood of the data under the new prior without requiring labeled data in the target domain. The procedure is applied independently in magnitude bins of width 0.5 mag as follows:

\begin{enumerate}
    \item \textit{Initialization:} We compute the empirical fraction of stars in the training sample, $\pi_{\rm train}$, for the given magnitude bin. The target prior is initialised to an arbitrary value, $\pi^{(0)}_{\rm full} = 0.5$.
    
    \item \textit{Expectation (E-step):} Using the current estimate of the target prior, $\pi^{(k)}_{\rm full}$, we update the individual probabilities for all $N$ sources in the bin using Eq.~\ref{eq:adjust-pstar-prob} to obtain $p_{{\rm cal}, i}^{(k)}$.
    
    \item \textit{Maximization (M-step):} We update the estimate of the target prior by averaging the calibrated probabilities over the full ensemble of sources in the bin:
    \begin{equation}
        \pi^{(k+1)}_{\rm full} = \frac{1}{N} \sum_{i=1}^{N} p_{{\rm cal}, i}^{(k)}
    \end{equation}
    
    \item \textit{Convergence:} Steps 2 and 3 are repeated until the estimate of $\pi_{\rm full}$ stabilises: $|\pi^{(k+1)} - \pi^{(k)}| < 10^{-6}$.
\end{enumerate}

\begin{figure}
\begin{center}
\includegraphics[width=\columnwidth]{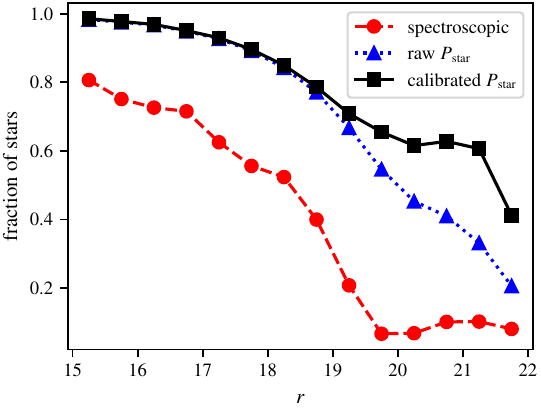}
\end{center}
\vspace{-0.4cm}
\caption[]{Estimated fraction of stars as a function of $r$-band magnitude, calculated in bins of 0.5 mag. Red circles represent the fraction of spectroscopically classified stars in the training sample. Blue triangles show the mean raw \pstar\ predicted by {\sc LeMoNNADE} for the full J-PLUS DR4 catalogue. Black squares indicate the mean calibrated \pstar\ for the full catalogue after applying the prior shift correction via the Expectation-Maximization algorithm. \label{fig:calibrate-pstar}}
\end{figure}

Figure~\ref{fig:calibrate-pstar} shows the impact of this calibration on the estimated stellar fraction as a function of magnitude. At bright magnitudes ($r \lesssim 19$), the calibration has a negligible effect because the raw predictions are highly confident (predominantly $p_{\rm raw} \approx 0$ or $1$). However, at fainter magnitudes ($r > 20$), where the raw model output tends to be ambiguous due to photometric errors and  biased by the scarcity of stars in the training set, the calibration significantly increases the estimated stellar fraction. 

\begin{figure}
\begin{center}
\includegraphics[width=\columnwidth]{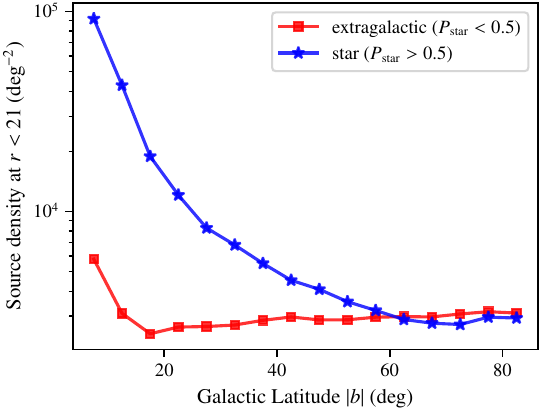} 
\end{center}
\vspace{-0.4cm}
\caption[]{Density of $r < 21$ stars and extragalactic sources as a function of Galactic latitude $|b|$ inferred from calibrated \pstar\ values.\label{fig:density-galat}}
\end{figure}

\begin{figure}
\begin{center}
\includegraphics[width=\columnwidth]{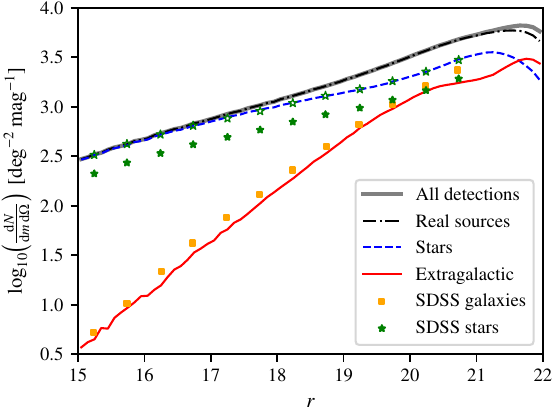}
\end{center}
\vspace{-0.4cm}
\caption[]{Differential number counts as a function of $r$-band magnitude. The thick grey solid line represents the distribution of all detections in the catalogue. The black dot-dashed line corresponds to the estimated population of real astrophysical sources, obtained by weighting each detection by $(1 - P_{\rm noise})$. The contributions from extragalactic sources (red solid line) and Galactic stars (blue dashed line) are derived by further weighting the real sources by $(1 - P_{\rm star})$ and $P_{\rm star}$, respectively. For comparison, we plot the number counts of stars (green solid stars) and galaxies (orange squares) from SDSS \citep{Yasuda2001}. The open green stars represent the SDSS stellar counts scaled by a factor of 1.5.\label{fig:counts-by-mag}}
\end{figure}

We test the robustness of the \pstar\ calibration by measuring the density of stars and extragalactic sources ($r < 21$) as a function of Galactic latitude, $b$ (Fig.~\ref{fig:density-galat}). As expected for an isotropic distribution, the density of extragalactic sources remains roughly flat, whereas the stellar density increases dramatically at lower Galactic latitudes. A subtle decrease in the extragalactic counts between $|b| = 80^{\circ}$ and $|b| = 20^{\circ}$ is consistent with higher average extinction at lower Galactic latitudes. At $|b| < 15^{\circ}$, we observe a dramatic increase in the apparent extragalactic counts; however, this excess can be entirely attributed to a minor contamination from misclassified stars, estimated at approximately 3.5\% for sources with $r < 21$.

We further validate the calibration of \pstar\ by comparing the number counts of J-PLUS stars and extragalactic sources with results from SDSS commissioning observations \citep{Yasuda2001}. 
To account for the classification uncertainty, we weight each J-PLUS source with its calibrated \pstar{} and (1 - \pstar) values, respectively. In addition, we use the probability of being a spurious detection, \pnoise, obtained in Appendix \ref{sec:remove-fakes}. 

Figure~\ref{fig:counts-by-mag} presents the differential number counts as a function of the $r$ magnitude for all J-PLUS sources detected (weight: 1), real J-PLUS sources (weight: 1 - \pnoise), stars (weight: (1 - \pnoise)\pstar), and extragalactic sources (weight: (1 - \pnoise)(1 - \pstar)).
Stellar counts are higher in J-PLUS compared to SDSS, but they agree remarkably well at all magnitudes if SDSS counts are scaled by a factor 1.5 (open stars in Fig.~\ref{fig:counts-by-mag}). The main reason for this discrepancy is that J-PLUS extends to lower Galactic latitude than SDSS, which increases the average counts for stars.
On the other hand, number counts for J-PLUS extragalactic sources appear below those of SDSS galaxies at all magnitudes. However, this is mostly a consequence of using $r$-band magnitudes uncorrected for Galactic extinction. Using the extinction-corrected $r$-band magnitude, extragalactic counts in J-PLUS agree with SDSS and other surveys (Fig. \ref{fig:counts-galaxies}).

\section{Probability of a false detection}
\label{sec:remove-fakes}

\begin{figure}
\begin{center}
\includegraphics[width=\columnwidth]{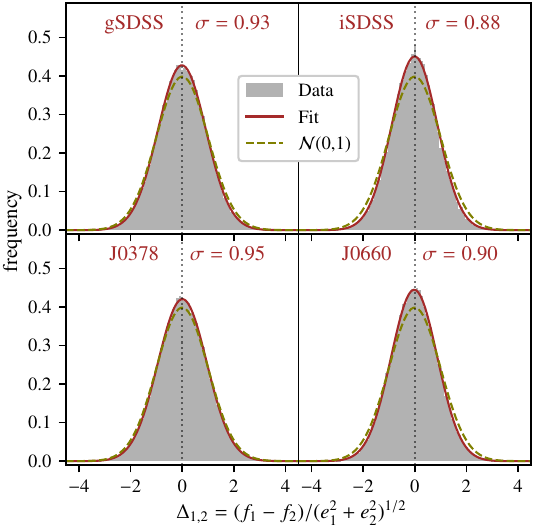}
\end{center}
\vspace{-0.4cm}
\caption[]{Distribution of the normalised flux difference \deltaonetwo\ between independent measurements of duplicated sources in four representative bands. The brown solid line shows the best-fitting Gaussian model while the olive dashed line represents the normal distribution expected for accurate nominal photometric errors.\label{fig:noise-model-dupes}}
\end{figure}

\begin{figure}
\begin{center}
\includegraphics[width=\columnwidth]{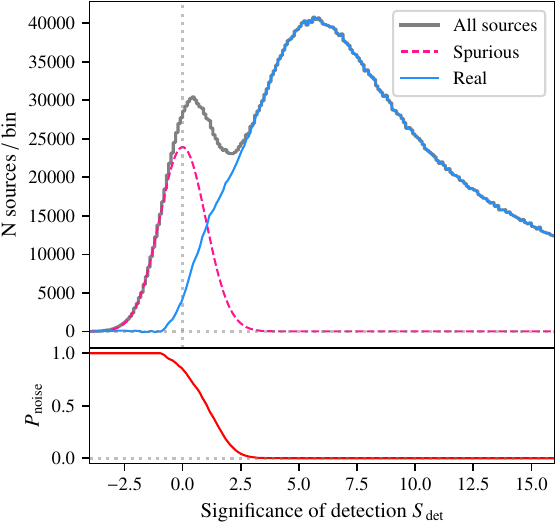}
\end{center}
\vspace{-0.4cm}
\caption[]{Top panel: Distribution of significance of detection in 11 J-PLUS bands excluding the detection band (\sdet). The thick grey line shows the total counts in bins of width 0.1. The pink dashed line represents the best-fitting model for the spurious source population. The thin blue line, obtained by subtracting the spurious model from the total counts, represents the predicted distribution for real sources. Bottom panel: Estimated frequency of false detections (\pnoise) as a function of the significance of detection.\label{fig:spurious-counts}}
\end{figure}

We estimated the probability that a given J-PLUS source corresponds to a spurious detection (noise artefact) rather than a real astrophysical object. We define this probability as \pnoise. The estimation of \pnoise\ relies on a statistical analysis of the signal consistency across the filter system, grounded in a validation of the photometric uncertainties.

First, we assessed the accuracy of the nominal photometric errors using sources detected in the overlapping regions between adjacent J-PLUS tiles. These duplicate detections provide two independent measurements of the same astrophysical source. We computed the normalised flux difference, \deltaonetwo, defined as:
\begin{equation}
    \Delta_{1,2} = \frac{f_1 - f_2}{\sqrt{e_1^2 + e_2^2}},
\end{equation}
where $f_1$ and $f_2$ are the flux measurements in the two tiles, and $e_1$ and $e_2$ are their respective nominal uncertainties. To ensure that the error budget is dominated by the sky background rather than calibration errors, we restricted this analysis to sources with $S/N$ $< 5$ in each band. Under the assumption of Gaussian errors and accurate uncertainty estimation, the distribution of \deltaonetwo\ should follow a standard normal distribution, $\mathcal{N}(0,1)$. However, as shown in Fig.~\ref{fig:noise-model-dupes}, the observed distributions are consistently narrower than the theoretical expectation. This indicates that the nominal photometric errors produced by the reduction pipeline are slightly overestimated. By fitting a Gaussian model to the \deltaonetwo\ distribution with the mean and standard deviation as free parameters, we derived correction factors for each band. We found that the nominal errors require a reduction of approximately 5--12\%, depending on the specific filter.

Building on these calibrated errors, we developed a metric to distinguish real sources from noise. For a spurious detection triggered by noise fluctuations in the detection band ($r$), the fluxes measured via forced photometry in the remaining bands have an expectation value of zero, with a variance determined solely by the sky background. We defined the combined detection significance, \sdet, as:
\begin{equation}
    S_{\rm det} = \frac{1}{\sqrt{N}} \sum_{i=1}^N \frac{f_i}{e_i \sigma_i},
\end{equation}
where the index $i$ runs over the $N$ J-PLUS bands excluding the detection band (i.e., $N=11$), and $\sigma_i$ represents the error correction factor derived from the duplicate analysis.

For a population of false detections, the distribution of \sdet\ is expected to follow $\mathcal{N}(0,1)$. The observed distribution of \sdet\ for all catalogued sources with $r<22$ is clearly bimodal (see Fig.~\ref{fig:spurious-counts}, top panel). It displays a peak near $S_{\rm det} \sim 0$, dominated by false detections, and a second peak at $S_{\rm det} \sim 6$, corresponding to the population of real sources.

We isolated the contribution of false detections by fitting the negative tail of the observed \sdet\ distribution (where the contribution from real sources is negligible) with a scaled $\mathcal{N}(0,1)$ model. The extrapolation of this model, shown as the pink dashed line in Fig.~\ref{fig:spurious-counts}, represents the distribution of \sdet\ for the spurious population. Subtracting this model from the total counts yields the distribution of real sources (blue solid line). Finally, we estimated the probability of a false detection, \pnoise, as a function of significance:
\begin{equation}
    P_{\rm noise}(S_{\rm det}) = N_{\rm false}(S_{\rm det}) / N_{\rm tot}(S_{\rm det}),
\end{equation}
where $N_{\rm false}$ is the number of spurious sources predicted by the noise model and $N_{\rm tot}$ is the total number of observed sources at a given \sdet.

This modelling indicates that approximately 7.5\% of all sources with $r<22$ are spurious. This contamination rises steeply at the faint end, reaching $\sim$25\% when considering only sources in the magnitude range $21.5 < r < 22$.

\section{Origin of excess of photometric redshift counts at $z > 0.6$}
\label{sec:highz-excess}

As noted in Section \ref{sec:z-distribution}, the inferred redshift distribution for J-PLUS sources exhibits an excess at $z > 0.6$ when compared to the SHELS survey. In this appendix, we investigate the potential origins of this discrepancy.

One possible cause for the excess counts in J-PLUS is uncertainty in the spectral classification derived from \textsc{LeMoNNADE}. The presence of sources with ambiguous classifications (i.e., \pstar\ values intermediate between 0 and 1) could bias the inferred redshift distribution, even if the \pstar\ values are statistically well-calibrated. Such a distortion would arise if the classification uncertainty is not uniform across all redshifts, for instance, increasing at redshifts where the observed colours of galaxies or quasars become degenerate with those of stars. Conversely, the photometric redshift solutions for stellar contaminants tend to cluster at these same degenerate redshifts.

We assess the impact of classification uncertainty on the inferred redshift distribution by using external morphological classifications. We selected all sources with $r < 23$ from the Hyper Suprime-Cam Subaru Strategic Program \citep[HSC-SSP;][]{Aihara2018,Aihara2022} Public Data Release 3 database within the region defined by $213^\circ < \rm{RA} < 249^\circ$ and $42^\circ < \rm{Dec} < 45^\circ$. This catalogue was cross-matched with the J-PLUS DR4 catalogue using a matching radius of $0.5\arcsec$, yielding 473,762 common sources (hereafter, the HSC+J-PLUS sample). We classify as bona-fide galaxies the HSC+J-PLUS sources for which the extendedness flags are set to 1 in the HSC $g$, $r$, and $i$ bands.

\begin{figure}
\begin{center}
\includegraphics[width=\columnwidth]{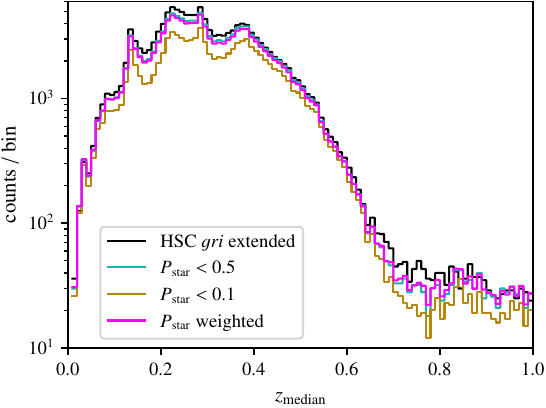}
\end{center}
\vspace{-0.4cm}
\caption{Photometric redshift distribution for $r<21$ galaxies in the JPSV field derived using different selection criteria. The black line represents bona-fide galaxies classified as extended in the HSC $g$, $r$, and $i$ bands. The cyan and gold lines correspond to J-PLUS sources selected with calibrated {\sc LeMoNNADE} thresholds of \pstar\ $< 0.5$ and \pstar\ $< 0.1$, respectively. The magenta line shows the distribution for all sources weighted by $w = 1 - P_{\rm{star}}$.\label{fig:zdistrib-jpsv}}
\end{figure}

\begin{figure*}
\begin{center}
\includegraphics[width=\textwidth]{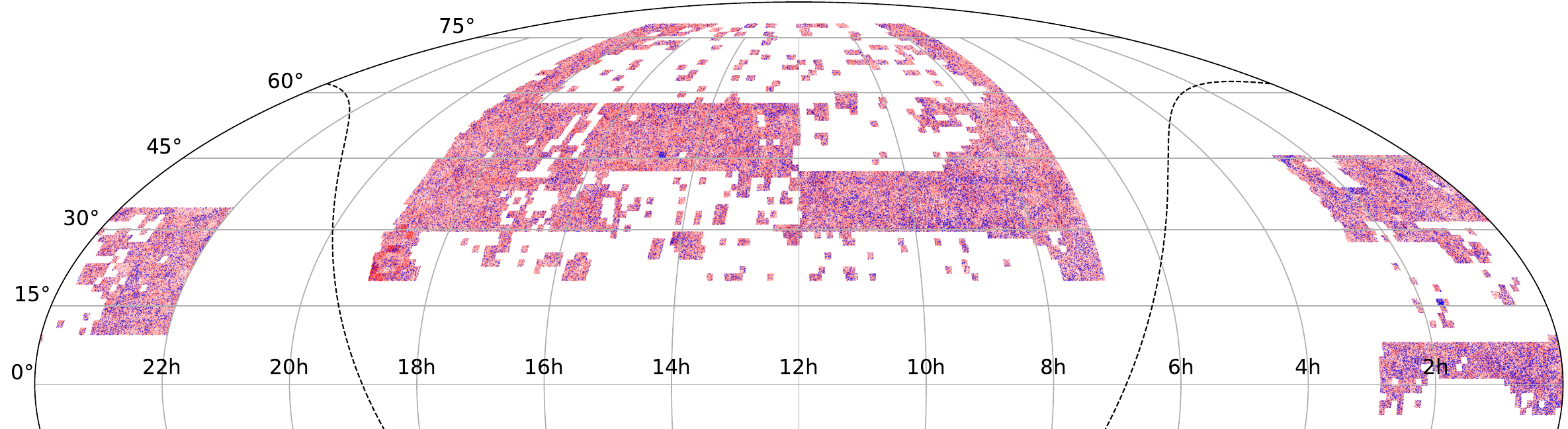}
\end{center}
\vspace{-0.4cm}
\caption{Spatial distribution across the J-PLUS footprint of all sources with $r < 21$ and \pstar\ $< 0.5$ (red dots) and the subset with $0.6 < z_{\rm{median}} < 1.0$ (blue dots). High overdensities of $z_{\rm{median}} > 0.6$ sources are visible at the coordinates of M31 (RA = 00h42m, Dec = +41$^\circ$16') and in tiles with atypically shallow broadband observations (e.g., RA = 14h40m, Dec = +45$^\circ$45'; RA = 01h41m, Dec = +15$^\circ$47'). The dotted line represents the Galactic plane.\label{fig:footprint}}
\end{figure*}

\begin{figure}
\begin{center}
\includegraphics[width=\columnwidth]{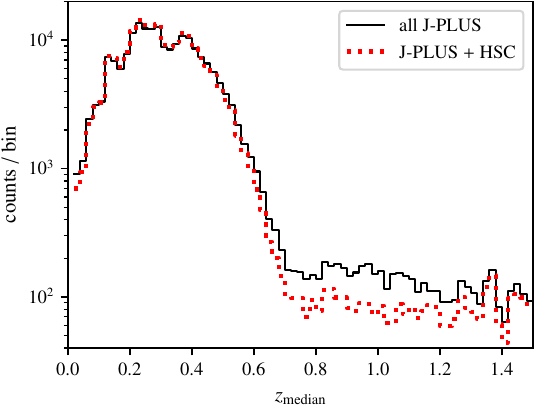}
\end{center}
\vspace{-0.4cm}
\caption{Photometric redshift distribution ($z_{\rm{median}}$) of $r < 21$ J-PLUS sources with \pstar\ $< 0.5$ in the JPSV field. The distribution of the full J-PLUS sample (black solid line) is compared to the subset of sources with an HSC counterpart within 0.5\arcsec\ (red dotted line), scaled by a factor of 1.522 to account for incomplete HSC spatial coverage.\label{fig:zdistrib-fakes}}
\end{figure}

\begin{figure}
\begin{center}
\includegraphics[width=\columnwidth]{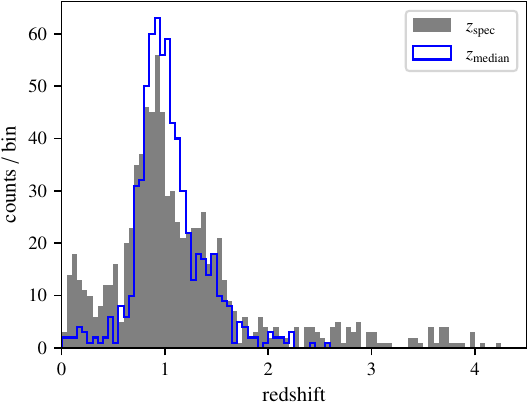}
\end{center}
\vspace{-0.4cm}
\caption{Distributions of spectroscopic redshift ($z_{\rm{spec}}$) and photometric redshift ($z_{\rm{median}}$) for sources in the \textsc{LeMoNNADE} spectroscopic test sample with low detection significance ($S_{\rm det} < 2$). Both distributions exhibit a prominent peak at $z \sim 1$.\label{fig:zdistrib-sdet2}}
\end{figure}

Figure~\ref{fig:zdistrib-jpsv} presents a comparison of the photometric redshift distributions for these bona-fide galaxies and for HSC+J-PLUS sources selected via different cuts in \pstar. We observe that the shape of the redshift distribution remains remarkably consistent across all selection criteria for the entire $0 < z < 1$ range. This suggests that no specific redshift interval is disproportionately affected by stellar contaminants or the misclassification of legitimate galaxies. The total number of sources classified as extended by HSC ($r < 21$) is approximately 8\% higher than the J-PLUS sample defined by $P_{\rm{star}} < 0.5$, and similarly higher than the total counts weighted by $(1 - P_{\rm{star}})$. This discrepancy indicates that the calibrated \pstar\ values may slightly overestimate the stellar probability, which is the opposite effect required to explain the excess of $z > 0.6$ sources relative to SHELS. Because the counts in the $0.6 < z < 1.0$ range are roughly equivalent for both morphological and \pstar-based classifications, we conclude that misclassified stars are not responsible for the excess counts at $z > 0.6$. The possibility that quasars contribute a significant fraction of the extragalactic counts in this range is also ruled out, as the counts remain consistent between morphologically extended sources and \pstar-selected extragalactic sources.

Further insight into the origin of these excess counts is provided by their spatial distribution. Figure~\ref{fig:footprint} shows the spatial distribution of J-PLUS sources with $r < 21$, \pstar\ $< 0.5$, and $0.6 < z_{\rm{median}} < 1.0$. While these sources cover the entire footprint, pronounced overdensities are present in tiles 196580 (RA = 14h40m, Dec = +45$^\circ$45') and 207151 (RA = 01h41m, Dec = +15$^\circ$47'). In both cases, the overdensity is bounded by the tile borders, indicating a data artefact rather than a physical source overdensity. These specific tiles have unusually shallow depths in the broad-band filters. Additionally, a strong overdensity of $0.6 < z_{\rm{median}} < 1.0$ sources occurs at the coordinates of M31 (RA = 00h42m, Dec = +41$^\circ$16'), resulting from \texttt{SExtractor} fragmenting the galactic disc into thousands of independent detections. This suggests that false detections and extraction artefacts contribute significantly to the $z > 0.6$ excess counts.
To verify it, Figure~\ref{fig:zdistrib-fakes} compares the $z_{\rm{median}}$ distribution of J-PLUS sources in the JPSV field with and without the requirement of an HSC counterpart within 0.5\arcsec. Given the incomplete HSC coverage in this field, the J-PLUS+HSC counts are scaled by a factor of 1.522 to match the overall source density. The two distributions agree remarkably well in the $0.05 < z_{\rm{median}} < 0.5$ range, but the J-PLUS+HSC counts are systematically lower at $z > 0.5$, diverging by $\sim$30--50\% in the $0.7 < z_{\rm{median}} < 1.2$ range. This confirms that the $z > 0.6$ excess is at least partially driven by spurious detections.

To understand why spurious sources are preferentially assigned photometric redshifts at $z > 0.6$, we examine the spectroscopic test sample used by \textsc{LeMoNNADE}. Figure~\ref{fig:zdistrib-sdet2} shows the distributions of $z_{\rm{spec}}$ and $z_{\rm{median}}$ for sources with low detection significance ($S_{\rm det} < 2$).
Notably, both distributions peak strongly at $z \sim 1$. We interpret this as a consequence of the DESI target selection function, which yields a redshift distribution centred at $z_{\rm{spec}} \sim 1$ for the faintest real sources in the training sample. Consequently, \textsc{LeMoNNADE} learns to output PDFs centred at $z \sim 1$ for any J-PLUS source exhibiting very low photometric $S/N$, including spurious detections. While the global rate of spurious detections at $r < 21$ is small, their concentration in anomalously shallow tiles, combined with their preferential assignment to the $z > 0.6$ range where real J-PLUS sources are sparse, can introduce significant systematics if \textsc{LeMoNNADE} photometric redshifts are used to analyse the high-redshift galaxy population.

\end{document}